\documentclass[aps,pra,two column,10pt]{revtex4-1}
\usepackage[english]{babel}
\usepackage{graphicx}
\usepackage{bbm}
\begin{document}

\title{On non-equilibrium photon distributions in the Casimir effect}
%
\author{Vanik E. Mkrtchian}
\affiliation{Institute for Physical Research, Armenian Academy of Sciences,
0203 Ashtarak-1, Republic of Armenia}
\author{Carsten Henkel}
\affiliation{Institute of Physics and Astronomie, University of Potsdam,
Karl-Liebknecht-Str.\ 24/25, 14476 Potsdam, Germany}
\begin{abstract}\noindent
The electromagnetic field in a typical geometry of the Casimir effect
is described in the Schwinger--Keldysh formalism.
The main result is the photon distribution function (Keldysh Green function)
in any stationary state of the field. 
A two-plate geometry with a sliding interface in local equilibrium 
is studied in detail, and full
agreement with the results of Rytov fluctuation electrodynamics is found.
As an illustration, plots are shown for a spectrum of the 
energy density between the plates.
\end{abstract}
\pacs{34.35.+a, 12.20.-m, 42.50.Nn}
\keywords{Casimir effect, Van der Waals interaction, quantum
friction, nonequilibrium electrodynamics of nanosystems.}

\maketitle


\makeatletter
\def\theequation{\thesection.\arabic{equation}}%
\renewcommand{\theequation}{\thesection.\arabic{equation}}%
\@addtoreset{equation}{section}
\makeatother

\maketitle

\section{Introduction}

In his seminal article 65 years ago, Casimir formulated a physical problem
\cite{Cas} 
which has had a tremendous influence on physics.
His pioneering analysis of the physical consequences of field
quantization under external, macroscopic boundary conditions 
is still of current relevance. Indeed, it turned out to be
one of the most prolific ideas in modern theoretical physics. 
Having a pure quantum background and being closely related to classical 
physics, the Casimir effect is a universal and wide-spread phenomenon
that can be found on all scales in the Universe.

One of the fascinating aspects of the Casimir effect is its simplicity, i.e.,
there are theoretical models that are amenable to exact solutions.
Leaving aside many important features of modern Casimir physics 
\cite{Dalvit_Book}, we consider the simple system of Ref.\cite{Cas}
with a slight generalization.
Namely, our setting consists of two half-spaces of
different materials with a plane-parallel gap in between. What
we know about the two media are their reflection coefficients (a matrix
in general) and macroscopic conditions like temperature distributions, 
macroscopic current densities, distance and relative motion parallel to 
the interfaces. We assume that these conditions are maintained 
stationary by external (generalized) forces.
We want to compute correlation
functions of the electromagnetic field in the gap. These yield for example
the average energy density of the field and the pressure exerted on the
two bodies as components of the stress (energy-momentum) tensor.
We focus in particular
on symmetrized correlations, also known as Keldysh-Green functions (KGF) 
in quantum kinetics, 
and are able to derive them under 
rather general circumstances in a stationary
Casimir geometry. We thus establish a natural non-equilibrium extension 
of Casimir's results.


Our approach starts from 
a simple observation which is a common feature among all
manifestations of the Casimir effect: the interaction between the two
interfaces disappears when their distance becomes large enough. 
On quite general grounds, we may therefore 
conclude that the electromagnetic field which we observe in
any Casimir system results from several fields that
originate in the material of the different half-spaces and in the vacuum
gap between them.
The distance dependence of the Casimir interaction is such 
that these fields become independent (uncorrelated) as the
interfaces are infinitely far away.
This fact is the cornerstone of our analysis: we use the large-distance limit
to express the KGF in a two-plate setting in terms of KGFs 
of independent half-spaces.

In this paper, we relax the assumption that the two plates share the same
temperature and state of motion. This makes it impossible to describe the
field between the plates as being in thermal equilibrium. 
The appropriate tool to calculate photon correlation functions is
then the Schwinger-Keldysh technique\cite{Schwinger_1961, Kel} of nonequilibrium
processes.
The application of the Keldysh formalism to the field of Casimir-Van der
Waals interactions has a relatively short history. The
first application, to our knowledge, is due to Janowicz, Reddig and
Holthaus \cite{Jan} in calculations of the electromagnetic heat
transfer between two bodies at different temperature. Sherkunov
used the non-equilibrium technique extensively for dispersion interactions
involving excited atoms and excited media \cite{Sher}. A general expression
for the electromagnetic force on an atom, in terms of the KGFs
of atom and radiation field, was found a few years ago
by one of the present authors \cite{vem}.

The physical processes 
behind these field-mediated interactions 
are the multiple reflections of photons between the
interfaces and their tunneling from one body to the other,
which follow from the boundary conditions for the electromagnetic field on the
interfaces.
To include these boundary conditions, we use an effective action
in the Schwinger-Keldysh technique with auxiliary fields and evaluate
generating functionals for KGFs by performing path integrals \cite{NO}.
Path integral approaches for the Casimir effect were introduced by Bordag,
Robaschik and Wieczorek \cite{Bord}. Li and Kardar considered the interaction 
between bodies mediated by a fluid with
long range correlations \cite{Meh2, Meh3}.
They applied a path integral technique to include
arbitrarily deformed bodies on which any kind of boundary condition can
be implemented. This feature makes the approach amenable to a perturbative
analysis of any deformed ideally conducting surfaces
\cite{Emig1, Emig2}. Further on, Emig
and B\"{u}scher \cite{Emig3} used the optical extinction theorem \cite{Born}
to reformulate the boundary conditions for a vacuum-dielectric interface
in  integral form that depends only on the fields on the vacuum side (i.e., in
the gap between two bodies). They then derived with path
integral techniques an effective Gaussian action for the photon gas in the
gap, providing the free energy and in particular the Casimir interaction
of dielectric bodies with arbitrary shaped surfaces. A similar approach has
been followed by Soltan et al.~\cite{Soltan} for a dispersive medium between the
bodies.
Recently, Behunin and Hu applied path integrals to problems of
the Casimir-Polder type (atom-surface interaction) in different
nonequilibrium 
%
situations: in Ref.\cite{Hu1} is considered the Van der Waals
interaction between an atom and a substrate in a stationary state out of
global equilibrium; Ref.\cite{Hu2} is calculating
atom-atom interactions in a quantized radiation field which is in a
nonequilibrium state.

Our theoretical approach to the non-equilibrium Casimir interaction 
where the boundary conditions involve different, locally defined
temperatures, is thus 
a synthesis between the Feynman path integral and the
Keldysh-Schwinger formalism of nonequilibrium processes \cite{KL, Alt}.

The paper is organized as follows: after preliminary notations and
introductions in Sec.\ref{s:2}, we introduce rather general expressions 
for retarded Green functions in the Casimir geometry of two planar
boundaries (Sec.\ref{s:3}). 
In Sec.\ref{s:4} is analyzed a Gaussian action in Schwinger-Keldysh space
which yields the non-equilibrium correlations for the fields (Keldysh-Green
functions). This action implements in particular the boundary conditions for
the retarded Green functions.
In Sec.\ref{s:5} we discuss
the retarded Green functions in the limit where the interfaces of the
Casimir system are infinitely removed from each other. 
Using this limiting
procedure, the Keldysh-Green function at any point in
the gap is expressed via Keldysh functions of the single interface systems
defined at the surfaces and via photon numbers in free space (Sec.\ref{s:7}). As an
application of the developed technique, we find in Sec.\ref{s:9} 
the field correlation functions in a Casimir geometry with a sliding interface.
This result is checked in
Appendix~\ref{a:sliding-reflection} where the same problem is solved using 
Rytov fluctuation electrodynamics.
%
We illustrate our results by analyzing the electromagnetic energy density 
between the plates in two typical non-equilibrium situations 
(Appendix~\ref{a:energy-spectrum}).
Concluding remarks are given in Sec.\ref{s:conclusions}, while technical
details are collected in the other Appendices.

\section{Preliminaries}
\label{s:2}

\subsection{Geometry of the problem}

We consider two bodies with 
parallel and homogeneous boundaries
located at $z = \pm a/2$. The boundaries are in stationary conditions: 
their temperatures are constant in time, and their relative motion 
(if any) is uniform and parallel to each other. 
We can then assume that the EM field in the cavity [$-a/2 \le z \le a/2$]
is stationary in time and homogenous in the $xy$-plane. As a consequence,
all relevant fields and correlation functions can be expanded in Fourier 
integrals with respect to frequency $\omega$ and wave vectors
${\bf q} = ( q_x, q_y )$ along the interfaces. 
We use in the following the shorthand 
\begin{equation}
	\Omega = (\omega , q_{x}, q_{y})
	\,.
	\label{eq:def-Omega}
\end{equation}
For fields like the vector potential, we get a mixed representation
\begin{equation}
	{\bf A}( \omega, q_x, q_y, z ) = {\bf A}( \Omega, z )
	\,,
	\label{eq:field-notation}
\end{equation}
where the argument $\Omega$ is suppressed where no confusion 
is possible. 
%

We work in the Dzyaloshinskii gauge $\varphi = 0$ where 
due to the transversality condition for the electric field, 
the normal component $A_{z}$ of the vector potential can be eliminated
in favor of the tangential ones $A_{x},\, A_{y}$. Furthermore, given the plane
of incidence spanned by ${\bf q}$ and the normal to the interfaces, 
the dynamical variables of the EM field in the cavity are the following
linear combinations
\begin{equation}
A_{s} = \frac{ q_{y}A_{x} - q_{x}A_{y} }{ | {\bf q}| }
\,,\qquad
A_{p} =  \frac{ {\bf q} \cdot {\bf A} }{ | {\bf q}| }
\,,
\label{eq:1.2}
\end{equation}%
which are nothing but the vector potential amplitudes of s- and p-polarized 
waves. In the following, we call the components defined in Eq.(\ref{eq:1.2}) 
the Weyl representation of the vector potential or we
say the vector potential is written in the Weyl basis \cite{Pieplow_2012}:
$\hat A = (A_s, A_p)$.

\subsection{Weyl representation of free space Green function}
\label{s:Weyl-free-space}

The Green function (GF) of the EM field plays a crucial role in the following.
Since it simply represents the vector potential due to a point current source,
it is also represented by a mixed Fourier representation 
involving a tensor
$D_{\alpha\alpha'}( \Omega ; z, z' )$ ($\alpha, \alpha' = x, y, z$).
The latter is the solution of
(given the gauge $\varphi = 0$) 
\cite{Landau_9}%
\begin{equation}
\left[ (\partial_z^{2} + q_{z}^{2})\delta_{\alpha\beta} - 
{\partial}_{\alpha } {\partial}_{\beta } \right]
D_{\beta \alpha'}( \Omega ; z, z' ) 
= 4\pi \delta_{\alpha\alpha'} \delta(z - z')
\label{eq:1.3}
\end{equation}%
where (we set $c = 1$)
\begin{equation}
q_{z}^{2} = \omega^{2} - q^{2}%
\,,\quad
{\partial}_{\alpha } = (iq_{x}, iq_{y},\partial_z )
\label{eq:1.4}
\end{equation}%
In the following, we only need the tangential part of the GF 
$D_{\sigma \sigma'}$ with $\sigma, \sigma' = x, y$. 
(See Eqs.(\ref{eq:3.15.a}--\ref{eq:3.15.c}) below for the normal components.)
Projecting
both indices into the Weyl basis according to Eq.(\ref{eq:1.2}),
we get from Eq.(\ref{eq:1.3}) the simple Helmholtz equation
\begin{equation}
(\partial_z^{2} + q_{z}^{2}) \hat{D}( \Omega ; z, z' ) 
= 4\pi \hat{g} \delta (z - z'),  
\label{eq:2.5}
\end{equation}%
where the $2\times 2$ matrix $\hat{g}$ reads 
\begin{equation}
\hat{g}=\left( 
\begin{array}{cc}
1 & 0 \\ 
0 & q_{z}^{2}/\omega^{2}%
\end{array}%
\right) 
\,.  
\label{eq:1.7}
\end{equation}

The GF in free space (no boundary conditions) is written
$\hat{ \Delta }( \Omega ; z, z' )$. As is well known, it comes
in two types: 
\textit{retarded} $\hat{\Delta}^{R}$ and \textit{%
advanced} $\hat{\Delta}^{A}$ Green functions. In the Weyl 
representation, they are given by
\begin{equation}
\hat{\Delta}^{R}( \Omega ; z, z' ) 
= \hat{\Delta}_{0} \, {\rm e}^{i q_{z} | z-z' | }  
= \hat{\Delta}^{A\ast }( \Omega ; z, z' )
\label{eq:1.8}
\end{equation}%
where%
\begin{equation}
\hat{\Delta}_{0} = \frac{2\pi }{iq_{z}}\hat{g}  
\label{eq:1.9}
\end{equation}%
and the wave vector $q_{z}$ is defined over the entire frequency axis by
\begin{equation}
q_{z} = \left\{ 
\begin{array}{ll}
\mathop{\rm sgn}\omega \sqrt{\omega^{2} - q^{2}} + i0
&\mbox{for}\quad \omega^{2} > q^{2}%
\\
i\sqrt{q^{2} - \omega^{2}}
&\mbox{for}\quad q^{2} > \omega^{2}
\end{array}%
\right.  \label{eq:1.10}
\end{equation}%
The two cases corresponding to outgoing \textit{propagating}
and to \textit{evanescent} waves, respectively.
The infinitesimal imaginary part of $q_{z}$ in ({\ref{eq:1.10}) secures the
analytical continuation of $\hat{\Delta}^R$ to complex frequencies 
in the upper half plane. 
It entails the existence of the limit%
\begin{equation}
\lim_{a \to +\infty} {\rm e}^{iq_{z}a} = 0  
\label{eq:1.11}
\end{equation}
for both propagating and evanescent waves. In addition, we have
the symmetry relation 
\begin{equation}
	q_z( - \Omega ) = - q_z^*( \Omega )
	\,.
	\label{eq:2.10}
\end{equation}
As a consequence, the retarded GF has the property
$\hat{ \Delta }^{R}( - \Omega ; z, z' ) = 
\hat{ \Delta }^{R\ast}( \Omega ; z, z' )$, 
as it should, being a response function between real-valued fields.

\section{Retarded Green functions}
\label{s:3}

\subsection{Single interface}

We include stationary and translation-invariant (in the $xy$-plane) 
boundary conditions in the retarded GF by adding reflected waves.
Let us start with a single interface at $z = - a/2$ where the GF (subscript
$-$) reads 
\begin{equation}
\hat{D}_{-}^{R}( \Omega ; z, z' ) = \hat{\Delta}^{R}( \Omega ; z, z' ) + 
\hat{R}_{-}\hat{\Delta}_{0} \,{\rm e}^{iq_{z}(z + z' + a)},  
\label{eq:3.14.a}
\end{equation}%
for $z, z' \ge - a/2$. The first term is the same as in free space.
The reflection matrix $\hat{R}_-$ at the lower
interface takes a simple diagonal form in the frame where the lower body is 
at rest
\begin{equation}
	\hat{R}_- = \left( \begin{array}{cc} 
		R^s_- & 0 \\ 
		0 & R^p_- 
	\end{array}
	\right)
	\label{eq:diagonal-R-matrix}
\end{equation}
whose matrix elements are for a
\begin{equation}
	\mbox{metal}: \quad
	R^s_- = \frac{ q_z \zeta( \omega ) - \omega 
			}{ q_z \zeta( \omega ) + \omega }
	\,,\quad
	R^p_- = \frac{ \omega \zeta( \omega ) - q_z  
			}{ \omega \zeta( \omega ) + q_z }
	\,,
	\label{eq:R-for-metal}
\end{equation}
where $\zeta( \omega )$ is the (dimensionless) impedance,
and for a 
\begin{equation}
	\mbox{dielectric}: \quad
	R^s_- = \frac{ q_z  - q_{z\varepsilon}  
			}{ q_z  + q_{z\varepsilon} }	
	\,,\quad
	R^p_- = \frac{ q_{z\varepsilon} - \varepsilon( \omega ) q_z
			}{ q_{z\varepsilon} + \varepsilon( \omega ) q_z }	
	\,,
	\label{eq:R-for-dielectric}
\end{equation}
where $\varepsilon( \omega )$ is the dielectric permittivity and
\begin{equation}
q_{z\varepsilon} = [ \varepsilon (\omega )\omega^{2} - q^{2} ]^{1/2}
\label{eq:2.4}
\end{equation}%
the wave vector in the lower medium.

If only the upper medium is present, we have a GF similar to 
Eq.(\ref{eq:3.14.a}), for $z, z' \le + a / 2$
\begin{equation}
\hat{D}_{+}^{R}( \Omega ; z, z' ) = 
\hat{\Delta}^{R}( \Omega ; z, z' ) + 
\hat{R}_{+} \hat{\Delta}_{0} \,{\rm e}^{-iq_{z} ( z + z' - a ) }
\,, 
\label{eq:3.14.b}
\end{equation}%
where $\hat{R}_+$ is the corresponding reflection matrix. 
We suppose here a generic form for $\hat{R}_{+}$.
The case of a sliding upper interface is discussed in 
Appendix~\ref{a:sliding-reflection}.

\subsection{General properties}

Let us collect a few general properties of the reflection matrices and the
GFs. It follows from Eq.(\ref{eq:d.14.a}--\ref{eq:d.14.d}) for $\hat{R}_{+}$ and
from the diagonal form
of $\hat{R}_{-}$ [Eq.(\ref{eq:diagonal-R-matrix})] that both fulfill the identity
\begin{equation}
\hat{g}\hat{R}_{\nu }^{T}=\hat{R}_{\nu} \hat{g}
\,,\quad \nu = \pm
  \label{eq:2.6}
\end{equation}%
The retarded GF~(\ref{eq:3.14.a}) therefore satisfies
\begin{equation}
\hat{D}^{R}( \Omega ; z, z' ) = [ \hat{D}^{R}( \Omega ; z', z ) ]^{T} 
\label{eq:2.8}
\end{equation}%
which is \textit{not} the reciprocity condition because the
wave vector $\mathbf{q}$ in the arguments of both sides is the same.

Taking into account Eqs.(\ref{eq:2.10}) and~(\ref{eq:d.14.a}--\ref{eq:d.14.d}),
the reflection matrices in Eq.(\ref{eq:diagonal-R-matrix}) and
Sec.\ref{a:sliding-reflection} satisfy the identity
\begin{equation}
\hat{R}_{\nu} (- \Omega ) = \hat{R}_{\nu }^{\ast} ( \Omega )  
\label{eq:2.11}
\end{equation}%
where the asterisk denotes the element-wise complex conjugation.
This entails that the symmetry relation is also valid for the retarded GF at 
a single interface
\begin{equation}
\hat{D}^{R}_\nu\left( -\Omega ;z,z' \right) =\hat{D}^{R\ast }_\nu\left(
\Omega ;z,z' \right)
\,, \quad \nu = \pm
\label{eq:2.9}
\end{equation}%
Below, we shall also deal with the advanced GF which
is defined as
\begin{equation}
	 \hat{D}^A_\nu( \Omega ; z, z' ) =
	 [ \hat{D}^R_\nu( \Omega ; z', z )]^\dag 
	 =
	 \hat{D}^{R\ast}_\nu( \Omega ; z, z' )
	\label{eq:def-advanced-GF}
\end{equation}
where the last equality follows from Eq.(\ref{eq:2.8}).

\subsection{Planar cavity}

The preceding properties carry over to the retarded and advanced GF
in the cavity formed by two interfaces. By adding up multiply reflected waves,
one finds the expression
\begin{equation}
\hat{D}^{R}( \Omega ; z, z' ) = \hat{\Delta}^{R}( \Omega ; z, z') + 
\sum\limits_{\nu ,\nu' =\pm} \hat{C}^{ \nu\nu' } 
\, {\rm e}^{iq_{z} ( \nu z + \nu' z' ) }  
\label{eq:3.1}
\end{equation}%
where
\begin{equation}
\hat{C}^{--} = \hat{U}_{+-}^{-1} \hat{R}_{+} \hat{\Delta}_{0}
\,{\rm e}^{iq_{z}a}
\,, \quad
\hat{C}^{-+} = \hat{U}_{+-}^{-1} \hat{R}_{+} \hat{R}_{-} 
\hat{\Delta}_{0}
\,{\rm e}^{2i q_{z}a}  
\label{eq:2.2}
\end{equation}%
and $\hat{C}^{++},\hat{C}^{+-}$ are defined by swapping indices
$-\leftrightarrow +$ in Eq.(\ref{eq:2.2}).
The matrix $\hat{U}_{+-}$ takes into account multiple reflections of 
photons in the cavity; it is given by the expression 
\begin{equation}
\hat{U}_{+-}=\hat{I}-\hat{R}_{+}\hat{R}_{-}\,{\rm e}^{2 i q_{z}a}
\,,
\label{eq:3.5}
\end{equation}%
where $\hat{I}$ is the unit matrix. An analogous formula gives
$\hat{U}_{-+}$. Eq.(\ref{eq:2.6}) above gives us the property%
\begin{equation}
\hat{U}_{-+}^{-1}\hat{g} = \hat{g} \hat{U}_{+-}^{-1T}  
\label{eq:2.7}
\end{equation}%
which ensures that the generalized reciprocity relation~(\ref{eq:2.8}) also
holds for the cavity GF~(\ref{eq:3.1}). Similarly, the 
relation~(\ref{eq:def-advanced-GF}) between retarded and
advanced GFs remains true as well.

\subsection{Normal components}

To conclude this Section, we give the tensor components involving
the normal direction. The following formulas can be shown from the
wave equation Eq. (\ref{eq:1.3}) and the symmetry properties above
($\lambda = s,p$ and $q =  |{\bf q}|$): 
\begin{eqnarray}
q_{z}^{2} D_{z\lambda }^{R} &=& i q\partial_z D_{p\lambda }^{R}
\,,  
\label{eq:3.15.a}
\\
q_{z}^{2}D_{\lambda z}^{R} &=& -i q\partial_{z'} D_{\lambda p}^{R}
\,, 
\label{eq:3.15.b}
\\ q_{z}^{2}D_{zz}^{R} &=& 
\frac{ q^2 }{ q_{z}^{2} } \partial_z \partial_{z'} D_{pp}^{R}
+ 4\pi \delta ( z - z' )
\,.  
\label{eq:3.15.c}
\end{eqnarray}%

\subsection{Generalized impedance matrices}

The reflection matrices $\hat{R}_\nu$ appearing in the expressions above
are the solutions to a scattering problem at the planar interfaces. We
discuss here an equivalent formulation in terms of generalized surface
impedances. These will provide the link between boundary conditions
imposed on fields and interactions with auxiliary fields restricted 
to the interfaces.

Evaluating the derivative with respect to $z$ and $z'$ of the retarded
GF~(\ref{eq:3.1}) at the interfaces $\pm a/2$, we come to the 
boundary conditions. With respect to the first coordinate $z$
(derivative $\partial_z$), we find
%
\begin{eqnarray}
&&
iq_{z}^{-1} \partial_z \hat{D}^{R}( -a/2, z' ) 
- \hat{Y}_{-} \hat{D}^{R}( -a/2, z' ) = 0  
\quad
\label{eq:4.1.a}
\\
&&
iq_{z}^{-1}\partial_z\hat{D}^{R}( +a/2, z')
+ \hat{Y}_{+} \hat{D}^{R}( +a/2, z' )=0  
	\label{eq:4.1.b}
\end{eqnarray}
where we defined the  matrices
\begin{equation}
\hat{Y}_{\nu} = (\hat{I} + \hat{R}_{\nu} )^{-1} (\hat{I} - \hat{R}_{\nu} )
\,,\quad
\nu =\pm  
	\label{eq:4.3}
\end{equation}
These generalize the concept of a surface admittance to a general
reflection problem. 
Indeed, from the reflection amplitudes~(\ref{eq:R-for-metal}) for a metallic
surface at rest, we get in the Weyl basis
\begin{equation}
%
	\mbox{metal:}\quad
	\hat{Y}_- = \frac{ 1 }{ \zeta( \omega ) }
	\left( \begin{array}{cc} 
	\omega / q_z & 0 \\ 0 & q_z / \omega
	\end{array} \right)
	\label{eq:admittance-metal}
\end{equation}
Note that despite multiple reflections, the boundary 
conditions~(\ref{eq:4.1.a}, \ref{eq:4.1.b})
are of local character: they link the fields and their normal derivatives 
at the same position with the corresponding admittance matrices.

With respect to the second coordinate $z'$ (derivative $\partial_{z'}$)
of the GF, a similar calculation yields
\begin{eqnarray}
&&
iq_{z}^{-1}\partial_{z'}\hat{D}^{R}( z, {-a/2} )
- \hat{D}^{R}( z, -a/2 ) \hat{Y}_{-}^{T} = 0  
	\label{eq:4.2.a}
\quad
\\
&&
iq_{z}^{-1}\partial_{z'}\hat{D}^{R}( z, {+a/2} ) 
+ \hat{D}^{R}( z, +a/2 ) \hat{Y}_{+}^{T} = 0  
	\label{eq:4.2.b}
\end{eqnarray}%
where the admittance matrices appear transposed. 

In the case of a single interface, we still find two boundary conditions
at $z, z' = \pm a/2$. If the ``missing'' body is the upper one, for example,
the admittance degenerates into $\hat{Y}_+ = \hat{I}$ 
from Eq.(\ref{eq:4.3}).
The boundary condition~(\ref{eq:4.1.b}) then becomes equivalent to
the Sommerfeld condition for an outgoing wave: 
$\hat{D}^R( z, z' ) \sim {\rm e}^{ i q_z z }$. This holds as long as
$z > z'$ and in both the propagating and evanescent sectors.

\subsection{Remarkable identity}

Using the boundary conditions~(\ref{eq:4.1.a}--\ref{eq:4.2.b}),
their counterparts for the advanced GF $\hat{D}^A$,
and the Green equation~(\ref{eq:2.5}), we come to the
following property of the GFs in the two-plate geometry
\begin{eqnarray}
\sum\limits_{\nu =\pm} 
	&& \hat{D}^{R}(z, \nu a/2 ) \hat{\Gamma}^{\nu } \hat{D}^{A}(\nu a/2, z') 
	\nonumber
	\\[-1ex]
	&& = 
	\hat{D}^{R}( z, z') - \hat{D}^{A}( z, z' )  
	\label{eq:4.7}
\end{eqnarray}
In Eq.(\ref{eq:4.7}) we have introduced effective source strengths
\begin{equation}
\hat{\Gamma}^{\nu } = 
- \frac12 \left(\hat{\Delta}_{0}^{-1}\hat{Y}_{\nu} - \mbox{h.c.} \right)  
	\label{eq:4.8}
\end{equation}%
which are easily shown to be antihermitian:
\(
\hat{\Gamma}^{\nu \dagger } = -\hat{\Gamma}^{\nu }  
\)
and symmetric:
\(
\hat{\Gamma}^{\nu T} = \hat{\Gamma}^{\nu }.  
\)
Therefore, they have purely imaginary matrix elements:
\(
\hat{\Gamma}^{\nu \ast }=-\hat{\Gamma}^{\nu }  
.
\)
Besides, using the parity properties~(\ref{eq:2.10}) and~(\ref{eq:2.11}) 
of $q_{z}$ and the reflection
matrices, the definition of the $\hat{\Gamma}^{\nu }$ entails
\begin{equation}
\hat{\Gamma}^{\nu }(-\Omega ) = \hat{\Gamma}^{\nu \ast}(\Omega )
= -\hat{\Gamma}^{\nu}( \Omega )
	\label{eq:4.9.d}
\end{equation}
The identity~(\ref{eq:4.7}) has a long history in macroscopic
fluctuation electrodynamics. It has been noted by Eckhardt
\cite{Eckhardt_1982}, although involving a spatial integral over volumes
where the imaginary part of the permittivity is nonzero. The
version we give here is technically somewhat simpler because only
surface sources appear. This may be related to the ``holographic principle''
stating that under certain circumstances, all relevant 
properties of a (source-free) field are encoded in a hypersurface.
This is obviously related to the classical Huyghens principle. An alternative
proof of Eq.(\ref{eq:4.7}) is given in
Appendix~\ref{a:remarkable-identity} using the Leontovich surface
impedance boundary condition.


If one of the two interfaces is missing, an identity similar to
Eq.(\ref{eq:4.7}) can be derived analogously. 
One simply has to replace the source strength
for the missing interface by
\begin{equation}
	\Gamma^\nu \mapsto \Gamma^0 = 
	- \frac12 \left(\hat{\Delta}_{0}^{-1}\hat{I} - \mbox{h.c.} \right)
	=
	- \hat{\Delta}_{0}^{-1} \Theta( \omega^2 - q^2 )
	\label{eq:4.13}
\end{equation}
where $\Theta$ is the unit step function. Note that only propagating
waves appear on the free boundary: they represent the fields incident from
infinity towards the interface.

\section{Non-equilibrium action}
\label{s:4}

We now address the key problem of this paper: evaluate correlations
for the EM field under non-equilibrium conditions. To this effect, we use
the path integral method and work with an action for the EM field.
An auxiliary field $\varphi$ is introduced to enforce the
boundary conditions at the two interfaces~\cite{NO}. This technique 
was developed in previous work \cite{Bord, Meh2, Meh3} for equilibrium
situations. 
We extend the approach to the whole Schwinger-Keldysh space and define 
a Gaussian action for two coupled Bose fields: the vector potential
$A$ and the auxiliary field $\varphi$
\begin{eqnarray}
S &=& \frac{1}{2}\int \big\{
\check{A}^{\dag} ( \Omega , z ) 
\check{\Delta}^{-1}( \Omega ; z, z' ) 
\check{A}( \Omega , z') 
+ \check{\varphi}^{\dag}( \Omega ) 
\check{F}\check{\varphi}( \Omega ) 
\nonumber
\\
&& + \check{\varphi}^{\dag}( \Omega ) 
\check{M}( \Omega, z ) \check{A}( \Omega , z ) 
+ \check{A}^{\dag}( \Omega, z ) 
\check{M}^{\dag}( \Omega, z ) \check{\varphi}( \Omega ) 
\big\}
\,.  
\nonumber
\\
\label{eq:6.1}
\end{eqnarray}
The integration is over $\Omega$ and $z$. In the action~(\ref{eq:6.1}),
the vector potential $\check{A}$ lives in Keldysh space and has two
components
that are called \emph{quantum} $\hat{A}^{\rm q}$ and 
\emph{classical} $\hat{A}^{\rm cl}$:
\begin{equation}
\check{A} = \left( 
\begin{array}{c}
\hat{A}^{\rm q} \\ 
\hat{A}^{\rm cl}%
\end{array}%
\right) 
\,,  \label{eq:6.3.a}
\end{equation}
each of which contains the familiar transverse amplitudes in the Weyl basis
\begin{equation}
\hat{A} = \left( 
\begin{array}{c}
A_{s} \\ 
A_{p}%
\end{array}%
\right) 
\,.  
\label{eq:6.3.b}
\end{equation}
The $4\times 4$ matrix $\check{\Delta}^{-1}$ is the inverse of the
Keldysh-Green (KG) matrix of the \emph{free} EM field. In the Keldysh 
basis~(\ref{eq:6.3.a}) for $\check{A}$, this KG matrix has the block
structure
\begin{equation}
\check{\Delta} = \left( 
\begin{array}{cc}
\hat{0} & \hat{\Delta}^{A} \\ 
\hat{\Delta}^{R} & \hat{\Delta}^{K}%
\end{array}%
\right)
\,,   
\label{eq:6.2}
\end{equation}%
where $\hat{\Delta}^{R,A}$ are the retarded and advanced Green functions
for free space, introduced in Sec.\ref{s:Weyl-free-space}. 
The function $\hat{\Delta}^{K}$ is the KGF for the free field, it collects
symmetrized correlations of the vector potential.
(A calculation is sketched in Sec.\ref{s:8} below.) 
Our goal is to calculate its counterpart 
in the presence of 
the two interfaces that we denote $\hat{D}^{K}$. It is given in the 
coordinate representation~\cite{X} as ($x = ( t, {\bf r} )$) 
\begin{equation}
D_{\alpha \alpha'}^{K}( x, x' ) = -i \left\langle 
		\{\hat{A}_{\alpha}( x ), \hat{A}_{\alpha'}( x' ) \}_+
	\right\rangle 
\label{eq:8.3}
\end{equation}%
where $\{ \cdot, \, \cdot \}_+$ is the anti-commutator.
We work here with the corresponding Fourier transforms with respect
to $t - t'$ and to the tangential coordinates. In the Weyl basis, this 
gives the matrix $\hat{D}^{K}( \Omega, z, z' )$.

The auxiliary field $\check{\varphi}$ in Eq.(\ref{eq:6.1}) has
eight components that are also grouped in the Keldysh structure
\begin{equation}
\check{\varphi} = \left( 
\begin{array}{c}
\hat{\varphi}^{\rm q} \\ 
\hat{\varphi}^{\rm cl}%
\end{array}%
\right) \,.  \label{eq:6.4.a}
\end{equation}%
The components $\hat{\varphi}^{\rm q, \,cl}$ themselves are
\begin{equation}
\hat{\varphi}=\left( 
\begin{array}{c}
\varphi _{-,s} \\ 
\varphi _{-,p} \\ 
\varphi _{+,s} \\ 
\varphi _{+,p}%
\end{array}%
\right) 
\,. 
\label{eq:6.4.b}
\end{equation}%
They are Weyl spinors localized in the lower 
(index $-$) and the upper ($+$) interface. 
Finally, the matrix $\check{M}( \Omega, z )$ 
in the action~(\ref{eq:6.1}) is related, as we shall see below, 
to the boundary conditions imposed at $z = \pm a/2$.

\subsection{Evaluating the path integral}

We follow the standard path integral procedure and
add source terms to the action~(\ref{eq:6.1})%
\begin{equation}
\frac{1}{2}\int (\check{J}^{\dag} \check{A} +\check{A}^{\dag} \check{J})
\end{equation}%
Then Gaussian path integrals are evaluated, first over the EM field and then 
over the auxiliary field. We get the generating functional for field correlations
whose expansion to second order in $\check{J}$ provides the following
expression for the KG matrix:
\begin{equation}
\check{D} = \check{\Delta} 
+ \check{\Delta}\check{M}^{\dag} \check{\Lambda} \check{M} \check{\Delta}  \label{eq:6.5.a}
\end{equation}
where $\check{\Lambda}$ is the solution of
\begin{equation}
\check{\Lambda}^{-1} = \check{F} - \check{M}\check{\Delta}\check{M}^{\dag}  
\label{eq:6.5.b}
\end{equation}
The KG matrix $\check{D}$ of Eq.(\ref{eq:6.5.a}) must 
have the block structure~(\ref{eq:6.2}) with off-diagonal 
blocks that are Hermitian conjugates 
[Eq.(\ref{eq:def-advanced-GF})]
and with an antihermitian Keldysh block%
\begin{equation}
[ \hat{D}^{K}(\Omega ; z, z' )]^{\dag} 
= -\hat{D}^{K}(\Omega ; z', z)
\,.  \label{eq:6.6}
\end{equation}
In addition, we impose the boundary conditions~(\ref{eq:4.1.a}--\ref{eq:4.2.b})
on the off-diagonal elements $\hat{D}^{R},\hat{D}^{A}$ of~(\ref{eq:6.5.a}).
These conditions unambiguously
define the structure of the matrices $\check{M}$ and $\check{F}$ 
in the action~(\ref{eq:6.1}) as 
\begin{equation}
\check{F}=\left( 
\begin{array}{cc}
\hat{F} & \hat{0} \\ 
\hat{0} & \hat{0}%
\end{array}%
\right)  
\label{eq:6.8}
\end{equation}%
where $\hat{F}$ is a $4\times 4$ antihermitian matrix, and 
\begin{equation}
\check{M} = \left( 
\begin{array}{cc}
\hat{0} & \hat{M} \\ 
\hat{M}^{\ast} & \hat{0}%
\end{array}%
\right)  \label{eq:6.7}
\end{equation}%
Here the matrix $\hat{M}$ collects the boundary conditions at the
two interfaces into a vector of operators in the Weyl basis 
\begin{equation}
\hat{M}(\Omega, z) = \left( 
\begin{array}{c}
\delta( z + a/2 ) [ iq_{z}^{-1}\hat{I}\partial_z - \hat{Y}_{-} ] \\ 
\delta( z - a/2 ) [ iq_{z}^{-1}\hat{I}\partial_z + \hat{Y}_{+} ]%
\end{array}%
\right)   
	\label{eq:4.5}
\end{equation}%
The boundary conditions~(\ref{eq:4.1.a}--\ref{eq:4.1.b}) then take the
integral form
(common argument $\Omega$ suppressed)
\begin{eqnarray}
0 &=& \int\! dz \, \hat{M}( z ) \hat{D}^{R}( z, z' )
	\label{eq:4.4.a}
\\
0 &=& \int\! dz' \, \hat{D}^{R}( z, z' ) \hat{M}^{T}( z' )
	\label{eq:4.4.b}
\end{eqnarray}%
with a derivative $\partial_{z'}$ acting to the left in the second line.
For the advanced GF, the boundary conditions~(\ref{eq:4.2.a}--\ref{eq:4.2.b})
can be written as integrals over 
$\hat{M}^{\ast}( z ) \hat{D}^{A}( z, z' )$
and over $\hat{D}^{A}( z, z' ) \hat{M}^{\dag}( z' )$.

With the choices~(\ref{eq:6.8}, \ref{eq:6.7}), one can show that the
Keldysh action~(\ref{eq:6.1}) acquires that so-called ``\textit{causal structure}" \cite{KL, Alt}.

Inserting the matrices $\check{M}$ and $\check{F}$ from
Eqs.(\ref{eq:6.7}, \ref{eq:6.8}) into Eq.(\ref{eq:6.5.a}), we find for the 
KG functions $\hat{D}^{R,A,K}$ the expressions
\begin{eqnarray}
\hat{D}^{R} &=& \hat{\Delta}^{R} + 
\hat{\Delta}^{R}\hat{M}^{T}\hat{\Lambda}\hat{M} \hat{\Delta}^{R}  
	\label{eq:6.9.a}
\\
\hat{D}^{A} &=& \hat{\Delta}^{A} + \hat{\Delta}^{A} \hat{M}^{\dag} 
\hat{\Lambda}^{\dag} \hat{M}^{\ast} \hat{\Delta}^{A}  
	\label{eq:6.9.b}
\\
\hat{D}^{K} &=& \big( \hat{I} + \hat{\Delta}^{R}\hat{M}^{T}
\hat{\Lambda}\hat{M} \big)
\hat{\Delta}^{K} \big( \hat{I} + \hat{M}^{\dag} \hat{\Lambda}^{\dag} 
\hat{M}^{\ast} \hat{\Delta}^{A} \big)
\nonumber
\\
&& {} + \hat{\Delta}^{R}\hat{M}^{T}\hat{\Lambda}\hat{F}
\hat{\Lambda}^{\dag} \hat{M}^{\ast} \hat{\Delta}^{A}
\label{eq:6.9.c}
\end{eqnarray}
where $\hat{\Lambda}$ is a $4\times 4$ matrix given by
\begin{equation}
\hat{\Lambda} = - (\hat{M} \hat{\Delta}^{R} \hat{M}^{T} )^{-1}  
\label{eq:6.10}
\end{equation}
%
The expressions~(\ref{eq:6.9.a}) and~(\ref{eq:6.9.b}) are integral forms of 
retarded and advanced GF because matrix products actually have to be read
as the concatenation of integral operators. It is trivial to check that they
satisfy the boundary conditions (\ref{eq:4.4.a}, \ref{eq:4.4.b}). 
Using the explicit form (\ref{eq:4.5}) of $\hat{M}$, we have checked in
a straightforward calculation
that Eq.(\ref{eq:6.9.a}) coincides with Eq.(\ref{eq:3.1}) for the retarded GF.

\subsection{Distribution (Keldysh-Green) functions for photons}

To analyse the expression~(\ref{eq:6.9.c}) for the KG function,
we first observe that
the first term is equal to zero because $\hat{\Delta}^{K}$ is a
solution of the homogeneous equation corresponding to Eq.(\ref{eq:2.5}). 
(See Appendix~\ref{a:homogeneous-DeltaK} for details.)
%
%
We shall argue that $\hat{F}$ that characterizes the auxiliary fields
can be taken in block-diagonal form
\begin{equation}
\hat{F}=\left( 
\begin{array}{cc}
4\hat{\Delta}_{0}\hat{P}(-)\hat{\Delta}_{0}^{\dag}  & \hat{0} 
\\ 
\hat{0} & 4\hat{\Delta}_{0}\hat{P}(+)\hat{\Delta}_{0}^{\dag} %
\end{array}%
\right)   
\label{eq:6.11}
\end{equation}%
where the quantity $\hat{P}( - )$ [$\hat{P}( + )$] correspond to the
lower [upper] boundary, respectively.
Adopting this choice, a tedious, but elementary calculation leads us to 
(common argument $\Omega$ suppressed)
\begin{equation}
\hat{D}^{K}( z, z' ) = 
\sum\limits_{\nu = \pm} 
\hat{D}^{R}( z, \nu a/2 ) \hat{P}( \nu ) \hat{D}^{A}( \nu a/2, z' )
	\,, 
\label{eq:6.12}
\end{equation}%
for the KG function. This has the same form as the Keldysh equation 
in quantum kinetics \cite{Kel}. If we would take this formal analogy serious,
then the quantities 
$\hat{P}(\Omega ;\nu )$ 
would coincide with the Keldysh polarization operators on the boundaries 
(i.e., the KG correlation of interface polarization fields). 
In the non-equilibrium theory, 
these are nonlinear functionals of the Keldysh function for
photons, so that in the semiclassical approximation, Eq.(\ref{eq:6.12}) 
would generate
a kinetic equation of Boltzmann type for the photon distribution function. 

In the two-plate geometry of the Casimir effect, however, this is not
the case because the
$\hat{P}( \nu )$ are \emph{independent} of the photon KG 
functions $\hat{D}^{R}$, $\hat{D}^{A}$, and $\hat{D}^{K}$. 
They are rather determined by given 
macroscopic states of the bodies and their interfaces and act like a 
linear driving for the EM field in the cavity. This suggests an interpretation 
of Eq.(\ref{eq:6.12}) in the spirit of Rytov's fluctuation electrodynamics:
the quantities $\hat{P}( \pm )$ encapsulate the distribution of
photon sources. They are the only piece of information needed to
produce the (non-equilibrium) distribution function
of photons in the cavity. For this reason, we call $\hat{P}( \pm )$
the \emph{photon sources} or \emph{fluctuating sources} in the following.

The above interpretation also helps to understand the 
diagonal form for the source distribution function
$\hat{F}$ in Eq.(\ref{eq:6.11}): the sources are 
located on the bodies' interfaces,
and correlations between the macroscopic states of 
the two bodies are neglected. We intend to investigate corrections to
this approximation in future work.

\section{The large distance limit}
\label{s:5}


In this Section, we consider the limit $a \to \infty$ in order to fix the strength
of photon sources by referring to the EM field in free space and near a single interface.


There are three different ways to take the limit:

({\bf A}) If we fix both points $z$, $z'$ in the cavity and go to the limit
$a \to \infty$, we recover Green functions in free space. In particular for
the retarded Green function,
\begin{equation}
      \lim\limits_{a \to \infty} \hat{D}^{R}( z, z' ) = \hat{\Delta}^{R}( z, z' )
      \label{eq:5.1a}
\end{equation}
with the free space Green function defined in Eq.(\ref{eq:1.8}).

({\bf B}) If the points stay at fixed distances $z, z' > 0$ from the lower interface,
we get 
\begin{equation}
      \lim\limits_{a \to \infty} \hat{D}^{R}( z - a/2, z' - a/2 ) 
      = \hat{D}^{R}_{-}( z, z'; a \to 0 ) 
      \equiv \hat{d}^{R}_{-}( z, z' )
      \label{eq:5.1b}
\end{equation}
where $\hat{D}^{R}_{-}$ is the Green function above a single interface. 
The notation $( \ldots ; a \to 0 )$ means that in the expression for
$\hat{D}^{R}_{-}$ [Eq.(\ref{eq:3.14.a})], 
$a$ should be set to zero.

({\bf C}) Similarly, we may take points at positions $z, z' < 0$ below the upper
interface and get the Green function below a single interface:
\begin{equation}
      \lim\limits_{a \to \infty} \hat{D}^{R}( z + a/2, z' + a/2 ) 
      = \hat{D}^{R}_{+}( z, z'; a \to 0 ) 
      \equiv \hat{d}^{R}_{+}( z, z' )
      \label{eq:5.1c}
\end{equation}
where the Green function $\hat{D}^{R}_{+}( z, z'; a \to 0 )$ is defined 
similar to case ({\bf B}) from Eq.(\ref{eq:3.14.b}).

We also need for the Keldysh-Green function Eq.(\ref{eq:6.12})
the limiting values when one position in the Green functions
recedes to infinity. Keeping $z$ fixed,
\begin{eqnarray}
	{a \to \infty}:\quad  && \hat{D}^{R}( z, \nu a/2 )
	\to 
	\hat{\Delta}^{R}( z, \nu ) \, {e}^{ i q_z a/2 }
	\,,
\\
	&& 
	\hat{\Delta}^{R}( z, \nu )
	=
	\hat{\Delta}_0 \, {e}^{ - i \nu q_{z} z }
	\label{eq:5.2.a}
\end{eqnarray}
In this limit, the exponential ${\rm e}^{ i q_z a/2 }$ in the first line
restricts the support 
to propagating waves (real $q_z$);  it drops out when products with
the advanced GF $\hat{D}^{A}$ are formed because of Eq.(\ref{eq:1.8}). 
In a similar way, we
define for the single-interface limits ({\bf B}, {\bf C}) the
functions
\begin{eqnarray}
	\hat{d}_{-}^{R}( z, \nu ) &=& 
	[ \hat{I} e^{ - i \nu q_{z} z } 
		+ \hat{R}_{-} e^{ i q_{z}z} ]
		\hat{\Delta}_{0}  
\label{eq:5.2.b}
\\
	\hat{d}_{+}^{R}( z, \nu ) &=&
	[ \hat{I} e^{ - i \nu q_{z} z}
		+ \hat{R}_{+} e^{ -i q_{z}z} ]
		\hat{\Delta}_{0}
\label{eq:5.2.c}
\end{eqnarray}

\section{Keldysh--Green functions}
\label{s:7}

We now consider the KG function derived at Eq.(\ref{eq:6.12}) in the 
limit $a\rightarrow +\infty$ in order to find the photon sources 
$\hat{P}( \nu )$.
Using the notation $\hat{\Delta}^{K}$, $\hat{d}_{\pm }^{K}$ 
and $\hat{P}_{0,-,+}$ in the three limits (\textbf{A}), ({\bf B}) 
and (\textbf{C}), we get
\begin{eqnarray}
\hat{\Delta}^{K}( z, z' ) &=& \sum\limits_{\nu} 
	\hat{\Delta}^{R}(z,\nu )
	\hat{P}_{0}(\nu )
	\hat{\Delta}^{A}(\nu ,z')
	\,,  
\label{eq:7.1.a}
\\
\hat{d}_{\mp}^{K}( z, z' ) &=& \sum\limits_{\nu }
	\hat{d}_{\mp}^{R}(z,\nu )
	\hat{P}_{\mp}(\nu )
	\hat{d}_{\mp}^{A}(\nu ,z')
	\,,  
\label{eq:7.1.b}
\end{eqnarray}%
where the GFs on the right hand sides are defined in 
Eqs.(\ref{eq:5.2.a}--\ref{eq:5.2.c}) above.
Manifestly, the sources in free space $\hat{P}_{0}(\nu )$ are defined by
the quantum state of \emph{free} EM field. Similarly, the fluctuation sources 
on the physical interfaces [the lower one, $\hat{P}_{-}(-)$ in case 
({\bf B}) and the upper one, $\hat{P}_{+}(+)$ in case ({\bf C})] 
are defined by the macroscopic quantum state of the corresponding bodies.

We assume the statistical independence of the bodies and the free 
EM field \cite{Henkel_2002}
and come to the conclusion that the sources on the `free interfaces'
coincide:
\begin{equation}
\hat{P}_{-}(+) = \hat{P}_{0}(+)
\, ; \quad
\hat{P}_{+}(-) = \hat{P}_{0}(-)
\,,
\label{eq:7.2.a}
\end{equation}%
and that the sources on the macroscopic interfaces are the same for one
and for two plates:
\begin{equation}
\hat{P}(-) = \hat{P}_{-}(-)
\,; \quad
\hat{P}(+) = \hat{P}_{+}(+)
\,.  
\label{eq:7.2.b}
\end{equation}
By evaluating Eq.(\ref{eq:7.1.b}) at $z = z' = 0$,
we can also express the interface sources in terms of the KG function 
there:
\begin{equation}
\hat{P}(\nu ) = \hat{\Delta}_{0}^{-1}
	(\hat{I} + \hat{R}_{\nu })^{-1} \hat{d}_{\nu}^{K}
	(\hat{I} + \hat{R}_{\nu }^{\dagger})^{-1}
	\hat{\Delta}_{0}^{-1\dagger}
	- \hat{P}_{0}( -\nu )  
\label{eq:7.3}
\end{equation}
where $\hat{d}_{\nu }^{K}$ is the boundary value of the KG function
for a system with a single interface
\begin{equation}
\hat{d}_{\nu }^{K} = \hat{d}_{\nu }^{K}(\Omega ; 0, 0)  
\label{eq:7.4}
\end{equation}
The free space photon sources $\hat{P}_0( \nu )$
are calculated in Appendix~\ref{s:8}.
In Sec.\ref{s:9}, we work out the quantities
$\hat{P}( \nu )$ for a specific example, allowing for the upper body
to be in uniform motion relative to the lower one.

We are now ready to collect our main result. The explicit expressions for 
the GFs $\hat{D}^{R, A}$ [Eqs.(\ref{eq:3.14.a},
\ref{eq:def-advanced-GF})] and for the photon sources 
$\hat{P}$ [Eq.(\ref{eq:7.2.b})]
are inserted into the KG function~(\ref{eq:6.12}) to give
\begin{equation}
\hat{D}^{K}( z, z' ) = 
	\sum\limits_{\nu, \nu' = \pm}
	\hat{D}^{\nu\nu'}
	e^{ i (\nu q_{z}z -\nu' q_{z}^{\ast} z') }
	\,. 
\label{eq:6.5}
\end{equation}%
We find the amplitudes
\begin{eqnarray}
\hat{D}^{--} &=& \hat{T}_{+} + e^{-2 {\rm Im}\, q_{z} a}
	\hat{R}_{+}\hat{T}_{-}\hat{R}_{+}^{\dagger }
	\,,
\label{eq:7.6.a}
\\
\hat{D}^{-+} &=& \hat{R}_{+} \hat{T}_{-} e^{iq_{z}a} + 
	\hat{T}_{+}\hat{R}_{-}^{\dagger } e^{-iq_{z}^{\ast }a}
	\,,
\label{eq:7.6.b}
\\
\hat{T}_{-} &=& \hat{U}_{-+}^{-1} \hat{\gamma}_{-}
	\hat{U}_{-+}^{-1\dagger}
	\,;
\label{eq:7.7.a}
\end{eqnarray}%
where $\hat{D}^{++}$, $\hat{D}^{+-}$, $T_{+}$ are found by swapping
the subscripts $+$ and $-$, $\hat{U}_{-+}$ has been defined in
%
Eq.(\ref{eq:3.5}), 
and 
\begin{equation}
\hat{\gamma}_{\nu } = e^{-{\rm Im}\, q_{z} a}
	(\hat{I}+\hat{R}_{\nu })\hat{\Delta}_{0}
	\hat{P}(\nu )\hat{\Delta}_{0}^{\dagger }
	(\hat{I}+\hat{R}_{\nu }^{\dagger})  
\label{eq:7.7.b}
\end{equation}%
These expressions give the distribution function of photons in a
planar cavity (homogeneous along the $xy$-directions and
stationary in time) whatever the macroscopic state of the two
boundaries, as encoded in $\hat{P}( \nu )$. Based on the assumption of
statistical independence, the $\hat{P}( \nu )$ are given by
reference situations with a single interface [Eqs.(\ref{eq:7.3}, \ref{eq:7.4})].

In some cases (for instance in the Casimir effect), one needs to subtract
the free-space KG function to get finite results for the relevant observables:
\begin{equation}
\hat{D}_{\rm ren}^{K} = \hat{D}^{K} - \hat{\Delta}^{K}  
\label{eq:7.8}
\end{equation}%
where $\hat{\Delta}^{K}$ is defined by Eq.(\ref{eq:7.1.a}) with 
free-space photon sources $\hat{P}_{0}( \nu )$ in the problem.

For completeness, the normal ($z$-) components of the KG functions are also
given here. They are expressed in terms of tangential components, using 
the homogenous version of Eq.(\ref{eq:1.3}). We get in analogy to 
Eqs.(\ref{eq:3.15.a}--\ref{eq:3.15.c}) for the retarded GF
\begin{eqnarray}
q_{z}^{2}D_{z\lambda }^{K} &=& i q\partial_z D_{p\lambda }^{K}
\,,  
\label{eq:7.9.a}
\\
q_{z}^{2}D_{\lambda z}^{K} &=& -i q\partial_{z'} D_{\lambda p}^{K}
\,, 
\label{eq:7.9.b}
\\
q_{z}^{2}D_{zz}^{K} &=& \frac{ q^{2} }{ q_{z}^{2}} 
	\partial_z \partial_{z'} D_{pp}^{K}
\,.
\label{eq:7.9.c}
\end{eqnarray}

\section{Example: sliding interfaces}
\label{s:9}

As an application of the theory developed so far, 
let us consider the KG function in the case of two bodies in relative
motion.
In this case in the limit ($\mathbf{C}$) we have
a sliding interface (moving parallel to $x$ with velocity $v$). 
We suppose that we have equilibrium in the body's rest frame (with 
%
temperature $T_{+}'$),
similar to Refs.\cite{Polevoi_1990, Kyasov_2002, Dedkov_2002b}.
Using the Lorentz covariant formulation
of the fluctuation-dissipation theorem \cite{Pieplow_2013}, we get for the 
KG function in the limit ($\mathbf{C}$) the expression
($k_B = \hbar = 1$)
\begin{eqnarray}
\hat{D}_{+}^{K}( \Omega, z, z' ) &=& [\hat{d}_{+}^{R}( \Omega, z, z' )
	- \hat{d}_{+}^{A}( \Omega, z, z' )]
	\coth \frac{ \omega' }{ 2T_{+}' }
\,;
\nonumber
\\
\\
\omega' &=& \gamma (\omega - v q_{x})  
\label{eq:9.1}
\end{eqnarray}%
where $\omega'$ is the Doppler-shifted frequency.
A calculation starting from Eq.(\ref{eq:7.3}) yields the corresponding 
source located at the upper surface
\begin{equation}
\hat{P}(+) = \hat{\Gamma}^{+} \coth \frac{ \omega' }{ 2T_{+}' }
\label{eq:P-plus}
\end{equation}%
where $\hat{\Gamma}^{+}$ is given in Eq.(\ref{eq:4.8}). The
reflection matrix $\hat{R}_+$ that appears in
$\hat{d}^{R}_{+}$ [see Eq.(\ref{eq:5.2.c})] 
is calculated in 
Appendix~\ref{a:sliding-reflection}, Eqs.(\ref{eq:d.13}--\ref{eq:d.15}).
%
These expressions cover any relative velocities $v$.

The lower interface is in equilibrium at $T_{-}$. Therefore%
\begin{equation}
\hat{D}_{-}^{K}( \Omega, z, z' ) = [ \hat{d}_{-}^{R}( \Omega, z, z' )
	- \hat{d}_{-}^{A}( \Omega, z, z' ) ]
	\coth \frac{ \omega }{ 2T_{-} }
\label{eq:9.3}
\end{equation}%
and then using Eq.(\ref{eq:7.3}) we get for the sources there
\begin{equation}
\hat{P}(-) = \hat{\Gamma}^{-} \coth \frac{ \omega }{ 2T_{-} }
\label{eq:c.4}
\end{equation}
Inserting 
Eqs.(\ref{eq:P-plus}, \ref{eq:c.4}) 
into Eq.(\ref{eq:6.12}),
we come to the KG function for the cavity with one sliding interface
\begin{eqnarray}
&& \hat{D}^{K}( \Omega; z,z' )
\label{eq:9.5}
\\
&& =
	\hat{D}^{R}( \Omega; z,-a/2)
	\hat{\Gamma}^{-}
	\hat{D}^{A}( \Omega; -a/2, z' )
	\coth \frac{ \omega }{ 2T_{-} }
\nonumber
\\
&& \quad {} + \hat{D}^{R}( \Omega; z, +a/2)
	\hat{\Gamma}^{+}
	\hat{D}^{A}( \Omega; +a/2, z' )
	\coth \frac{ \omega' }{ 2T_{+}' }
\nonumber
\end{eqnarray}
That the frequencies differ in the two $\coth$ terms has been known
in similar contexts since the pioneering work by Frank and Ginzburg 
on the Cherenkov effect (see 
Ref.\cite{Ginzburg_1996} for an overview).
This spoils any attempt to describe the sliding geometry by a global 
equilibrium assumption even if the two bodies are of the same temperature
(as in Ref.\cite{Philbin_2009}).
%
The sign change between $\omega$ and $\omega'$ (anomalous Doppler
effect) has also been noted in early work on field quantization in
moving media, see, e.g., Jauch and Watson \cite{Jauch_1948}.

\section{Conclusions}
\label{s:conclusions}

We have calculated the photon distribution function between two parallel plates
(Casimir geometry) under rather general conditions: 
on the two plates, any arbitrary stationary nonequilibrium state
is allowed for. Using 
the Schwinger-Keldysh formalism of nonequilibrium field theory, 
we could express the Keldysh-Green
function of photons at any point in the gap between the plates via the Keldysh 
functions of single interface systems,
defined at the surfaces, and via photon numbers in free space. As a cross check of 
the results, we consider one plate sliding relative to the other in local equilibrium,
and we find full coincidence of the results with Rytov theory,
%
without any restriction in relative velocities.

Our approach is flexible enough to allow for the zero-temperature
limit $T \to 0$ to be taken. The example of the sliding plates of Sec.\ref{s:9} 
then illustrates that one does \emph{not} recover an equilibrium situation.
The resulting frictional stress that opposes the relative motion is an interesting
%
(and controversial \cite{Philbin_2009, Dedkov_2010a, Pendry_2010, Barton_2011}) 
manifestation of 
an unstable vacuum state.
For similar situations, we may quote the Klein paradox \cite{IV}
(electron-positron pairs created by maintaining a 
%
static field
configuration), and
the Schwinger-Unruh effect (thermalization of an accelerated detector in
vacuum). Indeed, one possible explanation for quantum friction involves the
%
creation of particle pairs in the two plates \cite{Pendry_2010, Barton_2011, Maghrebi_2013b},
as pointed out in earlier work by Polevoi within the context of Rytov theory
\cite{Polevoi_1990}.

Finally, we suggest that the developed formalism is general enough to investigate
generalizations beyond the standard fluctuation electrodynamics. The crucial
assumptions of the latter are clearly spelled out: statistical independence
of the sources localized on the macroscopic bodies. If this is relaxed, one has
to establish the photon source strengths in some other way. Concepts from
non-equilibrium kinetic theory like the Boltzmann equation or the balance
of energy and entropy exchanges are likely to be instrumental here.

\appendix
\section{Remarkable surface identity}
\label{a:remarkable-identity}
%

In this Appendix, we give an alternative derivation of the remarkable
identity~(\ref{eq:4.7}) for the retarded and advanced Green functions. 
Coming back
to its interpretation as the electric field radiated by a point electric
dipole, we have the wave equation 
\begin{equation}
	\nabla \times ( \nabla \times {\bf E} ) - \omega^2 {\bf E} = 
	4\pi \omega^2 {\bf d} \, \delta( {\bf x} - {\bf x}' )
	\label{eq:wave-equation}
\end{equation}
where the source dipole ${\bf d}$ is located at ${\bf x}'$.
Multiply this equation by some vector field ${\bf F}$, to be specified later,
and integrate over a volume $V$ with boundary $A$. 
Performing a partial integration leads to
\begin{eqnarray}
	&& \int\!{\rm d}V \, ( \nabla \times {\bf F} ) \cdot ( \nabla \times {\bf E} )
	-
	\int\!{\rm d}A \,{\bf F} \cdot ( {\bf n} \times ( \nabla \times {\bf E} ) )
\nonumber \\
	&& {} - \omega^2 \int\!{\rm d}V \, {\bf F} \cdot {\bf E} =
	4\pi \omega^2 {\bf F}( {\bf x}' ) \cdot {\bf d}
	\label{eq:after-partial-int-1}
\end{eqnarray}
where ${\bf n}$ is the unit normal pointing \emph{into} the volume $V$. 
We choose for the volume $V$ the cavity bounded by two plates.
%
Use one of the Maxwell equations and apply
on the plates the
surface impedance boundary condition 
${\bf E}_t = \zeta {\bf n} \times {\bf H}$ 
due to Leontovich \cite{Landau_10},
where ${\bf E}_t$ is the tangential electric field.
%
This gives under the surface integral in Eq.(\ref{eq:after-partial-int-1})
\begin{eqnarray}
	&& {\bf F} \cdot ( {\bf n} \times ( \nabla \times {\bf E} ) )
	= {\rm i} \omega \, {\bf F} \cdot ( {\bf n} \times {\bf H} ) 
	= {\rm i}\frac{ \omega }{ \zeta } 
	\, {\bf F} \cdot {\bf E}_t 
	\qquad
	\label{eq:after-partial-int-2}
\end{eqnarray}
More general boundary conditions (as for dielectrics) could be included
by allowing for a ${\bf q}$-dependent impedance.
We now make the choice ${\bf F} = {\bf E}^*$ (complex conjugate) and 
take the imaginary part of Eq.(\ref{eq:after-partial-int-1}). This removes
the volume integrals
over the real functions $|{\bf E}|^2$ and $|\nabla \times {\bf E}|^2$. The
rest leads to
\begin{equation}
	\int\!{\rm d}A\,  {\rm Re}\left(\frac{ 1 }{ \zeta }\right)
	|{\bf E}_t|^2 
	= 4\pi \omega\, {\rm Im}
	\left( {\bf E}( {\bf x}' ) \cdot {\bf d}^* \right)
	\label{eq:remarkable-identity}
\end{equation}
We recognize on the right-hand side the imaginary part of the Green function,
${\rm Im}\, [ D^{R}_{ij}( \omega ; {\bf x}', {\bf x}' ) d_j d_i^* ]$ which
can be written as the difference between retarded and advanced GFs.
On the left-hand 
side, we recognize a source strength for surface currents given by
the (positive) real part
of the admittance $1/\zeta$. This integral indeed represents the
radiation by surface currents because the field
${E}_{ti}( {\bf x}) = 
D^{R}_{ij}( \omega; {\bf x}, {\bf x}' ) d_j$
is related, \emph{by reciprocity} 
%
[Eq.(\ref{eq:2.8})],
to the field generated at the vacuum point ${\bf x}'$ by a source at the 
boundary point ${\bf x}$.

Let us finally emphasize that Eq.(\ref{eq:remarkable-identity}) 
greatly simplifies calculations in fluctuation electrodynamics because
there is no need to perform a volume integration
over sources distributed throughout the bulk of the bodies. It is sufficient 
to specify the radiation generated by the bodies on their boundaries. 
A technique that can be applied with a similar advantage is the 
generalized Kirchhoff principle where so-called `mixed losses' are used 
to calculate correlation functions outside a body \cite{Dorofeyev_2011b}.

\section{Simplification of the KG function~(\ref{eq:6.9.c})}
\label{a:homogeneous-DeltaK}

We show here that the first line of Eq.(\ref{eq:6.9.c}) for
$\hat{D}^K$ vanishes:%
\begin{equation}
	0 =
\big( \hat{I} + \hat{\Delta}^{R}\hat{M}^{T}
\hat{\Lambda}\hat{M} \big)
\hat{\Delta}^{K} \big( \hat{I} + \hat{M}^{\dag} \hat{\Lambda}^{\dag} 
\hat{M}^{\ast} \hat{\Delta}^{A} \big)
	\label{eq:to-show}
\end{equation}
Recall that the KG function in free space 
$\hat{\Delta}^{K}( z, z' )$ solves the homogeneous equation
corresponding to Eq.(\ref{eq:2.5}). The dependence on the first
argument $z$ therefore reduces to ${\rm e}^{ {\rm i} \nu q_z z }$, 
$\nu = \pm$. We shall show that 
$-\hat{\Delta}^{R}\hat{M}^{T} \hat{\Lambda}\hat{M}$ acts like
the unit operator on these kind of functions and cancels with the first term
$\hat{I}$ in the left bracket of Eq.(\ref{eq:to-show}).

This program can be carried out by straightforward algebra, calculating
the matrix $\hat{\Lambda}$ by inversion from Eq.(\ref{eq:6.10}),
and working out the action of 
the boundary operator $\hat{M}$ [Eq.(\ref{eq:4.5})] on the exponentials:
\begin{equation}
	\int\!{\rm d}z\, \hat{M}( z ) \,{\rm e}^{ {\rm i} \nu q_z z }
	=
	\left( \begin{array}{c}
	{\rm e}^{ {\rm i} \nu q_z a/2 } [ - \nu \hat{I} - \hat{Y}_{-} ] 
	\\ 
	{\rm e}^{ - {\rm i} \nu q_z a/2 } [ - \nu \hat{I} + \hat{Y}_{+} ] 
	\end{array} \right)
	\label{eq:action-M-plane-wave}
\end{equation}

We present here a more compact proof whose starting point is
the product $\hat{\Delta}^{R}\hat{M}^{T}$, i.e., the line vector
\begin{eqnarray}
&& \int dz' \hat{\Delta}^{R}( z, z' )\hat{M}^{T}(z' )
\label{eq:B.3}
\\
&& =
e^{iq_{z}a/2} \left( 
\begin{array}{cc}
	e^{iq_{z}z} \hat{\Delta}_{0} (\hat{I} - \hat{Y}_{-}^{T})
	\,, & 
	-e^{-iq_{z}z} \hat{\Delta}_{0} (\hat{I} - \hat{Y}_{+}^{T})%
\end{array}%
\right)
\nonumber
\end{eqnarray}
By acting on this from the right with the block-diagonal, non-singular
matrix
\begin{equation}
	\hat{Q} = {\rm e}^{ - iq_{z}a/2 }
	\left( \begin{array}{cc}
	(\hat{I} - \hat{Y}_{-}^{T})^{-1} \hat{\Delta}_{0}^{-1} & 0
	\\
	0 & - (\hat{I} - \hat{Y}_{+}^{T})^{-1} \hat{\Delta}_{0}^{-1}
	\end{array} \right)
	\label{eq:def-helper-matrix}
\end{equation}
we get 
\begin{equation}
	\int dz' \hat{\Delta}^{R}( z, z' )\hat{M}^{T}(z' ) \hat{Q}
	=
	\left( 
\begin{array}{cc}
	e^{iq_{z}z} 
	\,, & 
	e^{-iq_{z}z} 
\end{array}
	\right)
	\label{eq:write-exps-as-operators}
\end{equation}
The action of the operator 
$\hat{\Delta}^{R}\hat{M}^{T} \hat{\Lambda}\hat{M}$
on these exponentials is precisely what we have to check. Using
the operator representation of Eq.(\ref{eq:write-exps-as-operators}), this
is easily worked out to be
\begin{eqnarray}
	&& \hat{\Delta}^{R}\hat{M}^{T} \hat{\Lambda}
	\hat{M}
	\left( 
\begin{array}{cc}
	e^{iq_{z}z} 
	\,, & 
	e^{-iq_{z}z} 
\end{array}
	\right)
\nonumber
\\	&& =
	\hat{\Delta}^{R}\hat{M}^{T} \hat{\Lambda}
	\hat{M} \hat{\Delta}^{R} \hat{M}^{T} 
	\hat{Q}	
\nonumber\\
	&& \stackrel{(*)}{=}
	- \hat{\Delta}^{R}\hat{M}^{T} \hat{\Lambda}
	\hat{\Lambda}^{-1}
	\hat{Q}
\nonumber\\
	&&	=
	- \hat{\Delta}^{R}\hat{M}^{T} \hat{Q}
	= 
	- \left( 
\begin{array}{cc}
	e^{iq_{z}z} 
	\,, & 
	e^{-iq_{z}z} 
\end{array}
	\right)
	\label{eq:trick-to-check}
\end{eqnarray}
where in step $(*)$ the definition Eq.(\ref{eq:6.10}) was used. 
The same cancellation with the unit operator $\hat{I}$ in
Eq.(\ref{eq:to-show}) happens for the operator in the right 
bracket there which is just hermitean conjugate to this one.

\section{Free-space sources}
\label{s:8}

To find the photon sources in free space $\hat{P}_{0}(\pm )$ required
for Eq.(\ref{eq:7.3}) we calculate the KG function by the standard mode 
expansion. The quantized vector potential is\cite{IV}
\begin{equation}
\mathbf{A}( x ) = \sum\limits_{\mathbf{k}\lambda}
\big( \hat{c}_{\mathbf{k}\lambda}
\mathbf{A}_{\mathbf{k}\lambda }
+ \hat{c}_{\mathbf{k}\lambda}^{\dagger}
\mathbf{A}_{\mathbf{k}\lambda }^{\ast})  
\label{eq:b.1}
\end{equation}%
where $\hat{c}_{\mathbf{k}\lambda}$ and 
$\hat{c}_{\mathbf{k}\lambda }^{\dagger }$ are the familiar
annihilation and creation operators for plane wave photon modes
(wave vector ${\bf k}$, polarization index $\lambda$).
The normalized mode functions are
($\hbar = c = 1$)
\begin{equation}
\mathbf{A}_{\mathbf{k}\lambda}( x ) = 
\sqrt{\frac{ 2\pi }{ \omega_{k}} }
\mathbf{e}_{\mathbf{k}\lambda }
e^{i (\mathbf{k} \cdot {\bf r} - \omega _{k}t)}
\,,
\quad
\omega_{k} = \left\vert \mathbf{k}\right\vert
\,.
   \label{eq:b.2}
\end{equation}%
This is inserted into the 
definition~(\ref{eq:8.3}) for the KG function 
$D_{\alpha \alpha'}^{K}( x, x')$ in the space-time domain.
%
We take the tangential components, switch to the Fourier-Weyl
representation and compare
with Eq.(\ref{eq:7.1.a}). In this way, we find for the photon sources in free 
space
\begin{eqnarray}
\hat{P}_{0}(\Omega ,\nu ) &=& \hat{\mathcal{N}}_{\nu }(\Omega )
	\hat{\Gamma}^{0} 
\label{eq:b.4.a}
\\
\hat{\mathcal{N}}_{\nu }(\Omega ) &=& 
\hat{I} \mathop{\rm sgn}\omega  + 
2[\Theta(\omega )\hat{N}_{\nu }(\Omega ) - 
(\Omega \rightarrow -\Omega )]  
\qquad
\label{eq:b.4.b}
\end{eqnarray}%
where we introduce two different matrices for left- and right-propagating
photons
\begin{equation}
\hat{N}_{\nu }(\Omega ) 
= \hat{N}(\mathbf{q},- \nu | q_{z} | )
\end{equation}%
which are, of course, diagonal in the polarization 
basis. They involve the Bose-Einstein distribution:
\begin{equation}
\hat{N} = \left( 
\begin{array}{cc}
N_{s} & 0 \\ 
0 & N_{p}%
\end{array}%
\right)
= \frac{ \hat{I}  }{ e^{ \omega / T } - 1 }
\label{eq:b.5}
\end{equation}
in the simple case of a uniform temperature $T$.
The photon sources in free space are thus defined by the average number 
of propagating photons that penetrate into the ``cavity"
through the corresponding ``surfaces".

\section{Rytov theory for a sliding interface}
\label{a:sliding-reflection}

For comparison with the non-equilibrium 
Keldysh-\-Schwin\-ger formalism,
we outline here a calculation for the two-plate system
within fluctuation electrodynamics, as developed by Rytov \cite{Ryt}.
%
More details can be found in Refs.\cite{Polevoi_1990, Dedkov_2010a, Maghrebi_2013b}.
For simplicity, we construct fluctuating sources based on surface currents
that are tangential to the two surfaces.
As a concrete example, we consider a ``sheared cavity'', i.e., the upper body
is in relative motion with velocity $v$ along the $x$-axis. Its
reflection matrix $\hat{R}_+$ is found by applying the Lorentz transformation.

\subsection{Spectral strength of surface currents}

We begin by considering the surface currents of the sliding
body in its rest frame $K'$. From electric charge conservation, we get the
charge density 
\begin{equation}
\rho' = \frac{ \mathbf{q}' \cdot \mathbf{j}' }{ \omega' }
\label{eq:d.1}
\end{equation}%
The currents in the laboratory frame $K$ are found with the help of
a Lorentz transformation. Using Eq.(\ref{eq:d.1}) for
the charge density $\rho'$, we find in the Weyl basis the representation
\begin{equation}
j_{\lambda }(\Omega, z) = O_{\lambda \lambda'}
j_{\lambda'}'(\Omega', z)
\,,\qquad
\Omega'\equiv (\omega', q_{x}', q_{y})  
\label{eq:d.2}
\end{equation}%
where the primed $k$-vector is given by the familiar 
Lorentz matrix (recall that $c = 1$)
\begin{equation}
\left( \begin{array}{c}
\omega' \\ q_{x}' 
\end{array} \right)
=
\left( \begin{array}{cc}
\gamma & - \gamma v \\
- \gamma v & \gamma
\end{array} \right)
\left( \begin{array}{c}
\omega \\ q_{x}
\end{array} \right)
\,,\qquad
\gamma =(1 - v^{2})^{-1/2}  
\,.
\label{eq:d.4.a}
\end{equation}%
For the matrix $\hat{O}$ in Eq.(\ref{eq:d.2}), the calculation yields
\begin{eqnarray}
%
\hat{O} &=& \frac{ \gamma }{ qq'\omega' }
\left( \begin{array}{cc}
\eta \omega' & -v q_{y} q_{z}^{2} \\ 
v q_{y} \omega \omega' & \eta \omega
\end{array} \right)   
\label{eq:d.3}
\\
%
q^{\prime 2} &=& q_{x}^{\prime 2} + q_{y}^{2}
\,,\qquad
\eta = q^{2} - v \omega q_{x}
\,.
\label{eq:d.4.b}
\end{eqnarray}

We assume that the current density contains only surface current
contributions $\hat{I}$
\begin{equation}
\hat{\jmath}(\Omega , z) = - \sum\limits_{\nu =\pm }
\hat{I}(\Omega ,\nu )
\delta (z-\nu a/2)  
\label{eq:d.5}
\end{equation}%
and get for the EM potential the source representation
\begin{equation}
{A}_{\lambda }(\Omega ,z) = 
\sum\limits_{\nu =\pm }
D_{\lambda \lambda'}^{R}(\Omega ; z, \nu a/2)
{I}_{\lambda'}(\Omega ,\nu )
\label{eq:d.6}
\end{equation}%
According to the framework of local Rytov theory\cite{Ryt}, 
we define the commutator of surface currents as follows
\begin{equation}
\left[ I_{\lambda }(\Omega, \nu ), \, I_{\lambda'}(\Omega', \nu') \right] 
= 
i \Gamma_{\lambda \lambda'}^{\nu}(\Omega )
\delta _{\nu \nu'}\delta (\Omega + \Omega') 
\label{eq:d.7}
\end{equation}
where the spectral strengths $\hat{\Gamma}^{\nu}(\Omega )$ 
($\nu = \pm$, matrices in the Weyl representation) are to be fixed.
The theory is local because currents ``living'' on different boundaries
commute and are un-correlated---this is the main assumption in Rytov theory.
To complete this definition, we calculate the commutator
of the vector potential $\hat A( \Omega, z )$ which is nothing but the 
retarded GF \cite{Landau_9}. Using the source representation~(\ref{eq:d.6})
and the source commutator~(\ref{eq:d.7}), we find
($\mathbf{q}$ omitted in all arguments)
\begin{eqnarray}
&& \hat{D}^{R}(\omega ;z,z') =
\nonumber\\
&&
\int\limits_{-\infty }^{+\infty }\!\frac{ d\omega _{1} }{ 2\pi }
\frac{
	\hat{D}^{R}(\omega _{1};z,\nu a/2)
	i\hat{\Gamma}^{\nu }(\omega _{1})
	\hat{D}^{A}(\omega _{1};\nu a/2,z')
	}{
	\omega -\omega _{1}+i0
	}  
\quad
\label{eq:d.8}
\end{eqnarray}%
Now using the Kramers-Kronig relations for the retarded GF, we come to 
\begin{eqnarray}
&& 2i\,{\rm Im}\,\hat{D}^{R}(\Omega; z, z') = 
\nonumber\\
&&
	\hat{D}^{R}(\Omega; z,\nu a/2)
	\hat{\Gamma}^{\nu }(\Omega )
	\hat{D}^{A}(\Omega; \nu a/2, z'). 
\label{eq:d.9}
\end{eqnarray}%
for the imaginary part. This equation coincides with (\ref{eq:4.7}) and
therefore the spectral strengths $\hat{\Gamma}^{\nu }$ must be given
by Eqs.(\ref{eq:4.8}). 

The definition~(\ref{eq:d.7}) of $\hat{\Gamma}^{\nu }$ as a commutator 
of the surface currents yields their transformation law
under a Lorentz transformation. 
Using Eq.(\ref{eq:d.2}), we find the rule%
\begin{equation}
\hat{\Gamma}^{+}( \Omega ) 
= \hat{O}\hat{\Gamma}^{+\prime}( \Omega' ) \hat{O}^{T}  
\label{eq:d.10}
\end{equation}
for the spectral strength on the upper (moving) interface ($\nu = +$).
The primed quantities are evaluated in the frame co-moving with the
body.

\subsection{Local equilibrium spectra}

We are now ready to compute the KG function according to 
its definition~(\ref{eq:8.3}), and have to consider equilibrium averages
of anticommutators for the surface currents. With the local equilibrium
assumption, these are computed in the rest frames of the interfaces.
For the lower interface, using the fluctuation--dissipation 
theorem~\cite{Landau_9} at temperature $T_{-}$, we get
\begin{eqnarray}
&& \left\langle \{ I_{\lambda }(\Omega_{1}, -) 
	,\, 
	I_{\lambda'}(\Omega_{2}, -)
	\}_{+}\right\rangle_{T_{-}}
	= 
\nonumber\\
&& 
	\coth\big(\frac{ \omega _{1} }{ 2T_{-} } \big)
	i\Gamma_{\lambda \lambda'}^{-}(\Omega_{1})
	\delta (\Omega_{1} + \Omega_{2})  
\label{eq:d.11.a}
\end{eqnarray}
For the upper interface in its rest frame $K'$, we have the same equation 
in terms of primed quantities, 
with the replacement $- \mapsto +$.
%
Using transformation laws~(\ref{eq:d.2}) for the currents 
and~(\ref{eq:d.10}) for $\hat{\Gamma}^{+}$, we get in the laboratory 
frame $K$
\begin{eqnarray}
&& \left\langle \{ I_{\lambda }(\Omega_{1}, +)
	,\,
	I_{\lambda'}(\Omega_{2}, +) \}_{+} \right\rangle_{T_{+}}
	=
\nonumber\\
&&
\coth \big(\frac{ \omega_{1}' }{ 2T_{+} }\big)
	i\Gamma_{\lambda \lambda'}^{+}(\Omega_{1})
	\delta(\Omega_{1} + \Omega_{2})  
\label{eq:d.11.b}
\end{eqnarray}
Inserting Eqs.(\ref{eq:d.11.a}, \ref{eq:d.11.b}) into the 
definition~(\ref{eq:8.3}) of the KG function, 
we come to the result~(\ref{eq:9.5}) found above within the 
Keldysh-Schwinger non-equilibrium framework.

\subsection{Lorentz-transformed reflection matrix}

From the transformation law~(\ref{eq:d.10}) 
for the surface current strength $\hat{\Gamma}^{+}$,
we can find the one for the reflection matrix $\hat{R}_{+}$
of the moving interface. Recalling Eq.(\ref{eq:1.9}) and the
fact that $q_z$ is an invariant for motions parallel to the
$xy$-plane, we get
\begin{equation}
\hat{g}^{-1}\hat{R}_{+}( \Omega ) = 
	\hat{O} \hat{g}^{\prime -1}
	\hat{R}'_{+}( \Omega' )
	\hat{O}^{T}  
\label{eq:d.13}
\end{equation}
In the co-moving frame $K'$, the matrix $\hat{R}'_{+}( \Omega' )$ is 
block-diagonal [see Eq.(\ref{eq:diagonal-R-matrix})] with elements
$R^{\prime s,p}_{+}( \Omega' )$. The transformation 
law~(\ref{eq:d.13}) yields the following Weyl components
\begin{eqnarray}
R_{+}^{ss}(\Omega ) &=& 
R_{+}^{\prime s}(\omega') \, \cos^2\theta
+ R_{+}^{\prime p}(\omega') \, \sin^2\theta
\,,  
\label{eq:d.14.a}
\\
R_{+}^{pp}(\Omega ) &=& 
R_{+}^{\prime p}(\omega') \, \cos^2\theta
+ R_{+}^{\prime s}(\omega') \, \sin^2\theta
\,,  
\label{eq:d.14.b}
\\
R_{+}^{sp}(\Omega ) &=& 
\frac{\omega}{q_{z}}
{} [ R_{+}^{\prime s}(\omega') - R_{+}^{\prime p}(\omega')]
\, \sin\theta \, \cos\theta
\,,  
\label{eq:d.14.c}
\\
R_{+}^{ps} &=& \frac{ q_{z}^{2} }{ \omega^{2}} R_{+}^{sp}
\,. 
\label{eq:d.14.d}
\end{eqnarray}%
Here, the polarization mixing is governed by the angle $\theta$ with
\begin{equation}
\sin\theta  = \frac{ v \gamma q_{y} q_{z} }{ q q' }\,.
\label{eq:d.15}
\end{equation}


















\section{Illustration: energy spectrum in the cavity}

This Appendix~E contains an expanded version of that one 
contained in the submission to
Ann.\ Phys.\ (Berlin). We give a number of technical details,
mainly as a cross-check for those who want to repeat these
calculations.

\subsection{Preparations}

\subsubsection{Energy density}

The average energy density $u(x)$ (in cgs units) is given by 
a sum of correlation functions
\begin{equation}
u( x ) = \frac{ 1 }{ 16\pi } 
\lim_{x' \to x}
\sum_{\alpha}
\langle \{ E_\alpha( x ), E_\alpha( x' ) \}_+ \rangle
+
\mbox{($E \mapsto B$)}
\label{eq:def-energy-density}
\end{equation}
where $\alpha = x, y, z$ is a cartesian index.
In the two-plate geometry considered in this paper, $u$ depends only 
on the $z$-coordinate and inherits from the KG function 
$\hat D^K( x, x' )$ the natural spectral representation 
$u( \Omega; z ) \,{\rm d}\omega \,{\rm d}^2 q / (2\pi)^3$ 
in terms of frequency $\omega$ and parallel wave vector
${\bf q} = ( q_x, q_y )$).

Calculating the electric and magnetic fields from the vector potential,
we find the following link to the KG function~(\ref{eq:8.3})
\begin{eqnarray}
&&
\langle \{ E_\alpha( x ), E_\alpha( x' ) \}_+ \rangle 
=
i \partial_t \partial_{t'} 
D^K_{\alpha\alpha}( x, x' ) 
\\
&&=
i \int\frac{ {\rm d}\Omega }{ (2\pi)^3 }
\omega^2 
D^K_{\alpha\alpha}( \Omega; z, z' ) 
\, {\rm e}^{ - i \omega ( t - t' )
+ i {\bf q} \cdot ({\bf x} - {\bf x}') }
\nonumber
\label{eq:uE-Fourier-expansion}
\end{eqnarray}
We abbreviate this link in the following by the notation
\begin{equation}
\langle \{ E_\alpha, E_\alpha \}_+ \rangle 
\to
i \omega^2 D^K_{\alpha\alpha}( \Omega; z, z' )
\label{eq:abbrev-link}
\end{equation}
where Eq.(\ref{eq:uE-Fourier-expansion}) permits us to identify
the rhs with the $\omega {\bf q}$-resolved 
spectral representation of the electric energy density. We eventually
put $z' = z$ and drop the arguments of $D^K_{\alpha\alpha}$ if no
confusion is possible.

The sum over the diagonal elements $D^K_{\alpha\alpha}( \Omega; z, z' )$ 
is reduced with the help of Eqs.(\ref{eq:7.9.a}--\ref{eq:7.9.c})
and the identities that follow from Eq.(\ref{eq:1.2})
\begin{eqnarray}
D^K_{x\alpha} &=& \frac{ q_x D^K_{p\alpha} + q_y D^K_{s\alpha} }{ q }
\,,
\nonumber
\\
D^K_{y\alpha} &=& \frac{ q_y D^K_{p\alpha} - q_x D^K_{s\alpha} }{ q }
\,,
\label{eq:}
\end{eqnarray}
the same equations holding for $D^K_{\alpha x}$ and $D^K_{\alpha y}$.
We find in terms of the components in the Weyl basis
\begin{eqnarray}
D^K_{xx} + D^K_{yy} &=&
D^K_{pp} + D^K_{ss}
\nonumber
\\
D^K_{zz} &=&
\frac{ q^2 }{ q_z^4 } \partial_z \partial_{z'}
D^K_{pp}
\nonumber
\\
\sum_{\alpha} D^K_{\alpha\alpha} &=&
\left( 1 + \frac{ q^2 }{ q_z^4 } \partial_z \partial_{z'}
\right)
D^K_{pp} + D^K_{ss}
\label{eq:}
\end{eqnarray}
A similar calculation for the magnetic energy density leads to
\begin{eqnarray}
\langle \{ B_x, B_x \}_+ + \{ B_y, B_y \}_+ \rangle &\to&
i 
\partial_z \partial_{z'} 
\left( D^K_{ss} + \frac{ \omega^4 }{ q_z^4 } D^K_{pp} \right)
\nonumber
\\
\langle \{ B_z, B_z \}_+ \rangle &\to&
i q^2 D^K_{ss} 
\label{eq:link-magnetic-density}
\end{eqnarray}
The energy spectrum is therefore determined by the
%
diagonal elements $D^K_{\sigma\sigma}( \Omega; z, z' )$,
i.e. the two transverse polarizations,
\begin{eqnarray}
u( \Omega; z ) &=& \frac{ i }{ 16\pi } 
\Big\{
( \omega^2 + q^2 + \partial_z \partial_{z'} ) D^K_{ss}
+ 
\nonumber\\
&&
{} + 
\omega^2 
\big( 1 + \frac{ \omega^2 + q^2 }{ q_z^4 } \partial_z \partial_{z'} 
\big)
D^K_{pp}
\Big\}
\label{eq:energy-spectrum-vs-KG}
\end{eqnarray}
where the derivatives $\partial_{z'}$ are evaluated at $z' = z$.

\subsubsection{Working out the Keldysh-Green function}

We use Eqs.(\ref{eq:6.5}--\ref{eq:7.7.b}) for the KG function. In the
sum over $\nu,\nu'$, the dependence on positions $z$ and $z'$ is made
explicit, and we can evaluate the derivatives in 
Eq.(\ref{eq:energy-spectrum-vs-KG}):
\begin{eqnarray}
u( \Omega; z ) &=& \frac{ i }{ 16\pi } 
\sum_{\nu\nu'}
\Big\{
( \omega^2 + q^2 + \nu \nu' |q_z|^2 ) D^{\nu\nu'}_{ss}
{\rm e}^{ i (\nu q_z - \nu' q_z^*) z }
\nonumber\\
&&
{} + 
\omega^2 
\big( 1 + \nu \nu' \frac{ (\omega^2 + q^2) |q_z^2| }{ q_z^4 }
\big)
D^{\nu\nu'}_{pp}
{\rm e}^{ i (\nu q_z - \nu' q_z^*) z }
\Big\}
\nonumber
\\
\label{eq:energy-as-nu-nuprime-sum}
\end{eqnarray}
The terms appearing here are given in Table~\ref{t:weighting-factors},
focusing on the two principal kind of waves: \emph{propagating} waves
with $q < \omega$ and real $q_z$ and \emph{evanescent} waves with
$q > \omega$ and imaginary $q_z$.

\begin{table*}[bht]
\begin{tabular}{lll}
	& propagating, $q_z = v \in \mathbbm{R}$ 
	& evanescent, $q_z = i \kappa$
\\
\hline
s-pol 
	& $[\omega^2 ( 1 + \nu\nu' ) + q^2 ( 1 - \nu\nu' )]
		e^{i (\nu - \nu') v z} $ 
	& $[\omega^2 ( 1 - \nu\nu' ) + q^2 ( 1 + \nu\nu' )]
		e^{- (\nu + \nu') \kappa z} $ 
\\
p-pol 
	& $(\omega/v)^2 [\omega^2 ( \nu\nu' + 1 ) + q^2 ( \nu\nu' - 1)]
			e^{i (\nu - \nu') v z} $
	& $(\omega/\kappa)^2 [\omega^2 ( \nu\nu' - 1) + q^2 ( \nu\nu' + 1)]
			e^{- (\nu + \nu') \kappa z} $
\end{tabular}
\caption[]{Weighting factors for energy density 
spectrum~(\ref{eq:energy-spectrum-vs-KG}), spelled out in polarization and 
type of waves.}
\label{t:weighting-factors}
\end{table*}
To work out the matrix elements $D^{\nu\nu'}_{\sigma\sigma}$, we start with
Eq.(\ref{eq:7.7.b}) for $\hat\gamma_\nu$. Using the surface current spectra
$\hat P( \nu )$ from Eqs.(\ref{eq:P-plus}, \ref{eq:c.4}) and the
abbreviations $\eta_+ = \coth \omega' / (2 T_+')$ and
$\eta_- = \coth \omega / (2 T_-)$, we get
\begin{eqnarray}
\hat{\gamma}_{\nu } &=& \frac{ -\eta_\nu }{ 2 } \, e^{-{\rm Im}\, q_{z} a}
	(\hat{I} - \hat{R}_{\nu})
	\hat{\Delta}_{0}^{\dagger}
	(\hat{I} + \hat{R}_{\nu}^{\dagger})  
	-
	{\rm h.c.}
\quad
\label{eq:}
\end{eqnarray}
where $\hat R_\nu$ is the reflection matrix for the interface at $z = \nu a/2$.
This formulation was used to generate the plots in 
Figs.\ref{fig:sliding-dielectrics}, \ref{fig:hot-n-cold} below.
In the following sections, we provide details on special cases where the
matrices $\hat R_\nu$ are diagonal (no relative motion). The matrices
$\hat\gamma_\nu$ are then diagonal as well and have elements
\begin{eqnarray}
	\gamma_{\nu,\sigma} 
	&=& \frac{ - 2\pi i g_{\sigma} \eta_\nu }{ v }
	(1 - |R_{\nu\sigma}|^2)
	\qquad \mbox{(prop.)}
\nonumber
\\
	&=& \frac{ - 4\pi i g_{\sigma} \eta_\nu \, e^{- \kappa a} }{ \kappa }
	{\rm Im}\, R_{\nu\sigma}
	\qquad \mbox{(evan.)}
\label{eq:gamma-prop-evan}
\end{eqnarray}
where $g_s = 1$ and $g_p = v^2 / \omega^2$ (resp., $- \kappa^2 / \omega^2$)
[see Eq.(\ref{eq:1.7})].
The notation for propagating and evanescent waves is the same as in
Table~\ref{t:weighting-factors}. Note in particular the Kirchhoff law
for propagating waves (emission and absorption are equal).


\subsection{Simple limiting cases}

\subsubsection{Free space}
\label{s:density-free-space}

The simplest reference situation is free space in global equilibrium 
at temperature $T$. The KG function 
$\hat{\Delta}^K$ 
is given by Eq.(\ref{eq:7.1.a})
with the source spectra $P_{0}( \nu )$~[Eq.(\ref{eq:b.4.a})]. We then
get 
\begin{equation}
u_0( \Omega; z ) = \frac{ \omega^2 }{ q_z } 
	\coth\big( \frac{ \omega }{ 2 T } \big)
\Theta( \omega^2 - q^2 )
\label{eq:free-space-energy-density}
\end{equation}
which is even in $\omega$ because of Eq.(\ref{eq:1.10}) defining $q_z$.
The step function reduces the spectral support to the light cone 
$q^2 + q_z^2 = \omega^2$ with real $q_z$. 

This result can be recovered from the general formalism by setting the
reflection matrices $\hat R_\nu \to 0$. The emission spectra
$\hat \gamma_\nu$ [Eq.(\ref{eq:gamma-prop-evan})] reduce to the 
propagating sector only. The other elements of Eqs.(\ref{eq:6.5}--\ref{eq:7.7.b})
become
\begin{eqnarray}
\hat T_\nu &=& \hat \gamma_\nu
\\
\hat D^{-+} &=& \hat D^{+-} = 0
\\
\hat D^{--} &=& \hat \gamma_+
\quad
\hat D^{++} = \hat \gamma_-
\label{eq:}
\end{eqnarray}
Since only the case $\nu = \nu'$ contributes, 
Eq.(\ref{eq:energy-as-nu-nuprime-sum}) yields a spatially constant
energy spectrum. Summing over the sources $\gamma_\pm$, we get from
Table~\ref{t:weighting-factors}:
\begin{equation}
u_0( \Omega ) 
= \frac{ 4 i }{ 16\pi }
2\omega^2 \frac{ - 2\pi i \eta }{ v } = 
\frac{ \omega^2 }{ v } \coth\frac{ \omega }{ 2 T }
\quad \mbox{(prop.)}
\label{eq:free-space-energy}
\end{equation}
which is nothing but Eq.(\ref{eq:free-space-energy-density}).

Summing over
positive and negative frequencies and integrating over ${\bf q}$ 
in the propagating sector, we get the Planck spectrum
\begin{equation}
w_0( \omega ) \frac{ d \omega }{ 2 \pi } 
= \frac{ \omega^3 \, d \omega }{ 2\pi^2 } \big\{
N( \omega ) + \frac12 \big\}
\label{eq:get-Planck}
\end{equation}
where $N( \omega )$ is the Bose-Einstein distribution and the $+1/2$
gives the zero-point energy.


\subsubsection{Symmetrical cavity, zero temperature}


As another check of Eq.(\ref{eq:energy-spectrum-vs-KG}), consider
a symmetrical cavity (identical plates at rest) at zero temperature. 
The energy density was calculated for this case by Sopova and 
Ford~\cite{Sopova_2002}. We first transform their integral representation
to real frequencies and then compare to our result, splitting into
propagating and evanescent waves.

Eq.(50) of Ref.\cite{Sopova_2002} for the energy density can be presented 
in the form
\begin{eqnarray}
u( z ) &=& \int\limits_{0}^{\infty}
\frac{\kappa^{3}\,d\kappa}{2\pi^{2}}
\int\limits_{0}^{1}dt\Big\{\frac{t^{2}R_{s}^{2}}{R_{s}^{2}-e^{2\kappa a}}
\\
&& {} +
\frac{(1-t^{2})R_{s}}{1-R_{s}^{2}e^{-2\kappa a}}e^{-\kappa a}\cosh 2\kappa z
+ (R_{s}\rightarrow - R_{p})
\Big\}
\nonumber
\end{eqnarray}
where we have adopted our conventions for the reflection coefficients (opposite 
sign in p-polarization) and for the location of the cavity boundaries. 
Here, $\kappa$ has the interpretation of a decay constant at imaginary
frequencies, and $t$ is related to the momentum parallel to the surfaces.
This can be made explicit with 
the change of variable $t = \xi / \kappa$, $1 - t^2 = q^2 / \kappa^2$:
\begin{eqnarray}
u(z) &=& 
\int\limits_{0}^{\infty}
\frac{d\xi }{2\pi^{2}}
\int\limits_{0}^{+\infty }dq
\frac{q}{\kappa }\Big\{\frac{-\xi ^{2}R_{s}^{2}e^{-2\kappa
a}+q^{2}R_{s}e^{-\kappa a}\cosh 2\kappa z
	}{ 1 - R_{s}^{2}e^{-2\kappa a}}
\nonumber
\\
&& {}
	+ (R_{s} \rightarrow -R_{p} )
\Big\}
\end{eqnarray}
Recalling the relation $\kappa^2 = \xi^2 + q^2$, this is actually 
the analytical continuation to imaginary frequencies 
$\omega = i \xi$ of a real-frequency integral. We shift the integration path 
to real frequencies where 
$i \kappa = i \sqrt{ q^2 - (\omega + i0)^2 } = q_z$, and read
off a frequency spectrum 
based on the measure $d\omega / 2\pi$:
\begin{eqnarray}
u( \omega; z ) 
&=& 
2 \mathop{\rm Re}
\int\limits_{0}^{\infty }\frac{q dq}{ 2\pi }
\Big\{
\frac{ \omega ^{2} R_{s}^{2} e^{2 i q_{z}a} 
+ q^{2} R_{s} e^{iq_{z}a}\cos 2q_{z}z
}{%
q_{z}(1 - R_{s}^{2} e^{2 i q_{z}a})
} 
\nonumber
\\
&& {}
+ (R_{s}\rightarrow - R_{p})
\Big\}
\end{eqnarray}%
The real part arises because the integral is made over positive frequencies only.
%
Splitting into propagating ($0 \le q \le \omega$) and evanescent waves
($\omega \le q$), we get
\begin{eqnarray}
&& 
u^{\rm pw}( \omega; z ) =
\int\limits_0^{\omega}\frac{ q dq }{ 2\pi } 
\Big\{
\frac{ 2\omega^2 
[ \mathop{\rm Re} ( R_s^2 e^{ 2 i v a } ) - |R_s|^4 ]
}{ v | 1 - R_s^2 e^{ 2 i v a } |^2 }
\nonumber
\\
&& {}
+
\frac{ 2 q^2 
\cos(2 v z ) (1 - |R_s|^2) \mathop{\rm Re} (R_s e^{ i v a })
}{ v | 1 - R_s^2 e^{ 2 i v a } |^2 }
\nonumber
\\
&& {} + (R_{s}\rightarrow - R_{p})
\Big\}
\label{eq:Sopova-Ford-prop}
\\
&&
u^{\rm ew}( \omega; z ) =
\int\limits_{\omega}^{\infty}\frac{ q dq }{ 2\pi } 
\Big\{
\frac{ \omega^2 e^{ - 2\kappa a } \mathop{\rm Im} R_s^2 
}{ \kappa | 1 - R_s^2 e^{ - 2 \kappa a } |^2 }
\nonumber
\\
&& {} +
\frac{ 2 q^2 \cosh( 2 \kappa z ) e^{ - \kappa a }
(1 + |R_s|^2 e^{ - 2 \kappa a })
\mathop{\rm Im} R_s 
}{ \kappa | 1 - R_s^2 e^{ - 2 \kappa a } |^2 }
\nonumber
\\
&& {} + (R_{s}\rightarrow - R_{p})
\Big\}
\label{eq:Sopova-Ford-evan}
\end{eqnarray}

Let us now come back to our approach. In Eqs.(\ref{eq:6.5}--\ref{eq:7.7.b}),
all matrices are diagonal and commute. We write $U_\sigma$ for the diagonal
elements of $\hat U_{-+} = \hat U_{+-}$ [Eq.(\ref{eq:3.5})].
In addition, both interfaces have the
same temperature so that $\eta_{\pm} = \eta$ and $\hat T_{\pm} = \hat T$ 
are the same. 
To simplify some of the following
expressions, we treat propagating and evanescent waves separately.
We recover each of two lines in Eqs.(\ref{eq:Sopova-Ford-prop},
\ref{eq:Sopova-Ford-evan}) individually from the KG function.

\paragraph*{Propagating waves.}
Here $q_z = v$ is real, and the terms with $\hat D^{++}$ and 
$\hat D^{--}$ contribute with the same weighting 
factor [Table~\ref{t:weighting-factors}]. Sum the two:
\begin{eqnarray}
D^{--}_{\sigma} + D^{++}_{\sigma} &=& 
2 T_{\sigma} (1 + |R_{\sigma}|^2 )
\nonumber\\
&=& 
\frac{ -4\pi i g_{\sigma} \eta }{ v |U_\sigma|^2 }
 (1 - |R_{\sigma}|^2) (1 + |R_{\sigma}|^2 )
\nonumber
\end{eqnarray}
At this point, we recall that Sopova and Ford~\cite{Sopova_2002} work with
a regularized energy density where the free-space value is subtracted.
This value can be found from Sec.\ref{s:density-free-space}, so we subtract
\begin{eqnarray}
(D^{--}_{\sigma} + D^{++}_{\sigma})_{0} &=& 
\frac{ -4 \pi i g_{\sigma} \eta }{ v }
\label{eq:}
\\
&=&
\frac{ -4 \pi i g_{\sigma} \eta }{ v |U_\sigma|^2 }
\left[ 
1 - 2 \,{\rm Re}\, ( R_{\sigma}^2 e^{ 2 {\rm i} v a } )
+ |R_{\sigma}|^4
\right]
\nonumber
\end{eqnarray}
See how this changes one prefactor and adds one term:
\begin{eqnarray}
(D^{--}_{\sigma} + D^{++}_{\sigma})_{\rm reg} &=& 
\frac{ -4 \pi i g_{\sigma} \eta }{ v |f_\sigma|^2 }
[ 2 \,{\rm Re}\, ( R_{\sigma}^2 e^{ 2 {\rm i} v a } )
- 2 |R_{\sigma}|^4 ]
\nonumber
\\
\end{eqnarray}
Multiplying with the coefficient $i/16\pi$ from Eq.(\ref{eq:energy-as-nu-nuprime-sum})
and the weighting factors from Table~\ref{t:weighting-factors}, we get
the following contribution to the energy density
\begin{equation}
\sum_{\sigma = {s,p}}
\frac{ \omega^2 \eta }{ v |U_\sigma|^2 }
[ {\rm Re}\, ( R_{\sigma}^2 e^{ 2 {\rm i} v a } )
- |R_{\sigma}|^4 ]
\label{eq:prop-constant-contrib}
\end{equation}
We get a spectrum over positive frequencies by adding the value at $-\omega$.
From the definition of the KG function, we have
\begin{equation}
\hat{D}^{K}\left( -\Omega ;z,z^{\prime }\right) = -[\hat{D}^{K}\left( \Omega
;z,z^{\prime }\right) ]^{\ast }
\end{equation}
so that we simply have to take twice the real part of 
Eq.(\ref{eq:prop-constant-contrib}).
This still needs to be integrated over ${\bf q}$, yielding
$\int_0^{\omega} q dq / (2\pi)$ in the propagating sector since nothing
depends on the orientation of ${\bf q}$.  
In this way,
we recover the first line of $u^{\rm pw}( \omega; z )$ 
in Eq.(\ref{eq:Sopova-Ford-prop}). ($\eta = \mathop{\rm sign} \omega$ 
at zero temperature.)

The terms involving $\hat D^{+-}$ and $\hat D^{-+}$ are handled in a 
similar way and give position-dependent contributions.
From Eq.(\ref{eq:7.6.b}), we get
\begin{eqnarray}
D^{-+}_{\sigma} e^{ - 2 i v z } &=& 
[ R_{\sigma} T_{\sigma} \, e^{ i v a }
+ R_{\sigma}^* T_{\sigma} \, e^{ - i v a }
] e^{ - 2 i v z }
\nonumber
\\
&=& \frac{ - 4\pi i g_{\sigma} \eta }{ v |U_\sigma|^2 }
(1 - |R_{\sigma}|^2) 
\mathop{\rm Re}( 
R_{\sigma} \, e^{ i v a }
)
e^{ - 2 i v z }
\nonumber
\end{eqnarray}
Adding the term $D^{+-}_{\sigma} e^{ 2 i v z }$ yields a cosine. 
There is no free-space term to subtract here. The weight factors
from Table~\ref{t:weighting-factors} now lead to different signs
for s- and p-polarization. Putting everything together, we find
from the previous expression the second line of $u^{\rm pw}( \omega; z )$ 
in Eq.(\ref{eq:Sopova-Ford-prop}).

\paragraph*{Evanescent waves.}
We begin now with the terms $\nu = -\nu'$ which do not depend on
$z$ [see Table~\ref{t:weighting-factors}]. Their sum is worked out as
\begin{eqnarray}
D_\sigma^{-+} + D_\sigma^{+-} &=&
\frac{ - 16 \pi i g_\sigma \eta \, e^{ - 2\kappa a } 
	}{ \kappa | U_\sigma |^2 } 
\mathop{\rm Re} R_{\sigma} \mathop{\rm Im} R_{\sigma} 
\nonumber
\end{eqnarray}
where the last two factors can also be written 
as $\frac12 \mathop{\rm Im} R_{\sigma}^2$.
We sum over the polarizations and get a positive-frequency spectrum
given by the first line of Eq.(\ref{eq:Sopova-Ford-evan}).

The final contribution starts with
\begin{eqnarray}
D_\sigma^{--} e^{ 2 \kappa z } &=& T_{\sigma} 
( 1 + e^{ - 2 \kappa a } |R_{\sigma}|^2 )
e^{ 2 \kappa z }
\nonumber\\
&=&
\frac{ - 4\pi i g_{\sigma} \eta \, e^{- \kappa a } 
}{ 
\kappa |U_\sigma|^2 
}
{\rm Im}\, R_{\sigma}
( 1 + e^{ - 2 \kappa a } |R_{\sigma}|^2 )
e^{ 2 \kappa z }
\nonumber
\end{eqnarray}
Adding $D_\sigma^{++} e^{ - 2 \kappa z }$ gives a
hyperbolic cosine. The polarization sum is actually a difference,
and we finally get the second line of Eq.(\ref{eq:Sopova-Ford-evan})
for $u^{\rm pw}( \omega; z )$.

\subsection{Two non-equilibrium examples}

\newcommand{\figpath}{/Users/carstenh/Work/Dropbox/Keldysh-Cavity/Revision/Figs/}

\begin{figure*}[htb]
\includegraphics*[width=75mm]{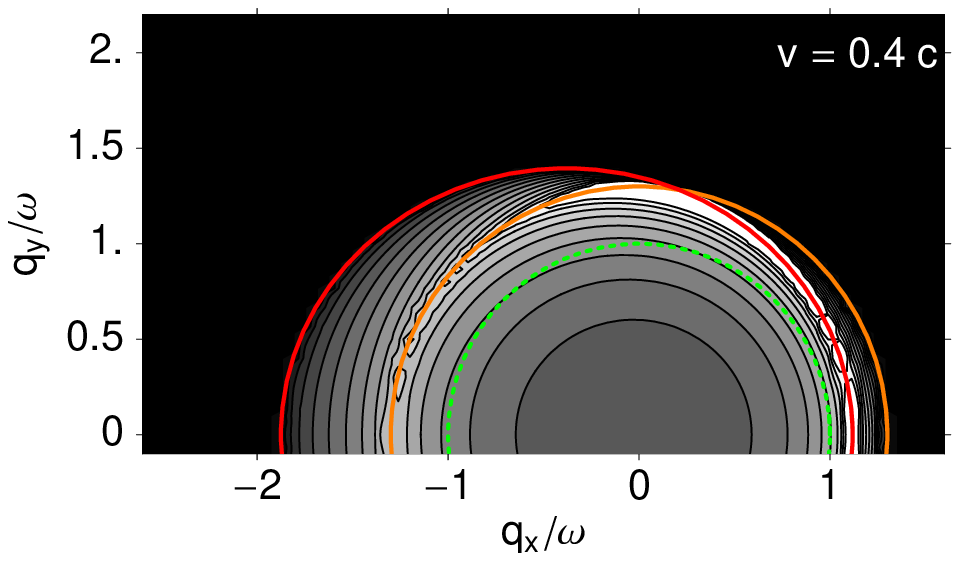}
\hspace*{3mm}
\includegraphics*[width=75mm]{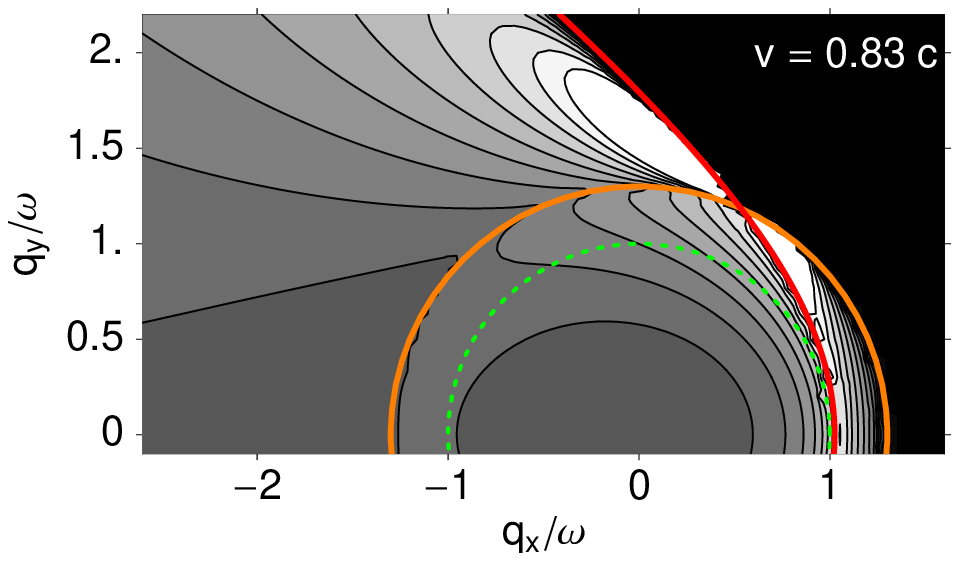}
\caption[]{%
Spectrum of electromagnetic energy per area $U( \omega, {\bf q} )$ between
two dielectric bodies, one moving at velocity $v$ along the $x$-axis,
at a fixed frequency $\omega$.
Parameters: zero temperature,
refractive index $n = 1.3$ (for both bodies in their respective 
rest frames, frequency-independent for simplicity),
distance $a = 0.4 / \omega \approx 0.064$ free-space wavelengths. 
The contours give the dimensionless quantity $U( \omega, {\bf q} )$ in steps
of $0.1$ between $0$ and $1.3$, higher values are clipped (white area), same
scale in both panels.
Dotted Green circle (radius $q = \omega$): 
free-space light cone; orange circle (radius $q = n \omega$): 
propagating photons in the dielectric at rest (`polariton cone'). 
Red ellipse (hyperbola): polariton cone in moving dielectric, as seen from the
laboratory frame. 
(\emph{left panel}) Velocity below Cherenkov threshold $v_c = c/n$: the polariton
cone is an ellipse. 
(\emph{right panel}) $v$ above Cherenkov threshold: the polaritons of the moving
dielectric fill two
hyperboloids. A non-zero energy spectrum, but with less structure, 
is found on the
other hyperboloid at $q_x > q_{x1} \approx 5.9\,\omega$ [Eq.(\ref{eq:ellipse-points})].
}
\label{fig:sliding-dielectrics}
\end{figure*}

\begin{figure*}[htbp]
\includegraphics*[height=60mm]{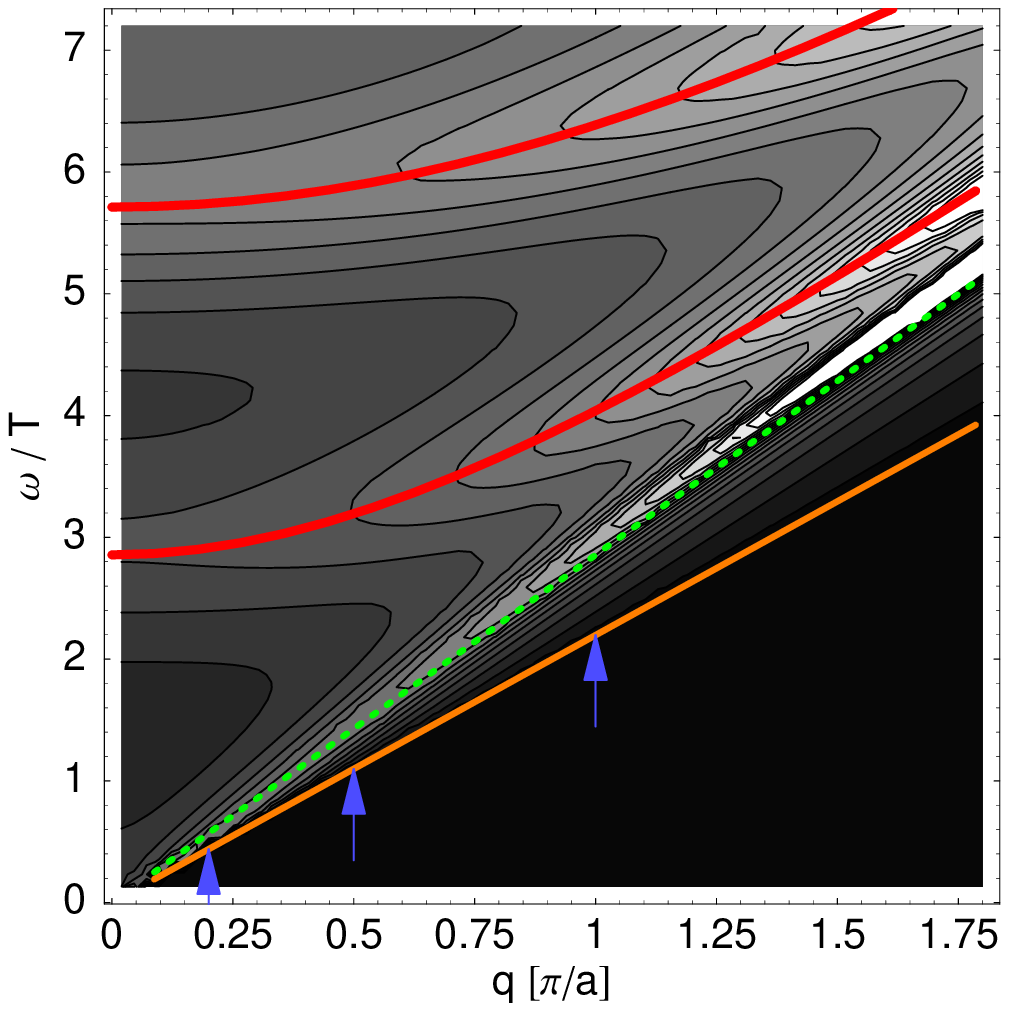}
\hspace*{3mm}
\includegraphics*[height=50mm]{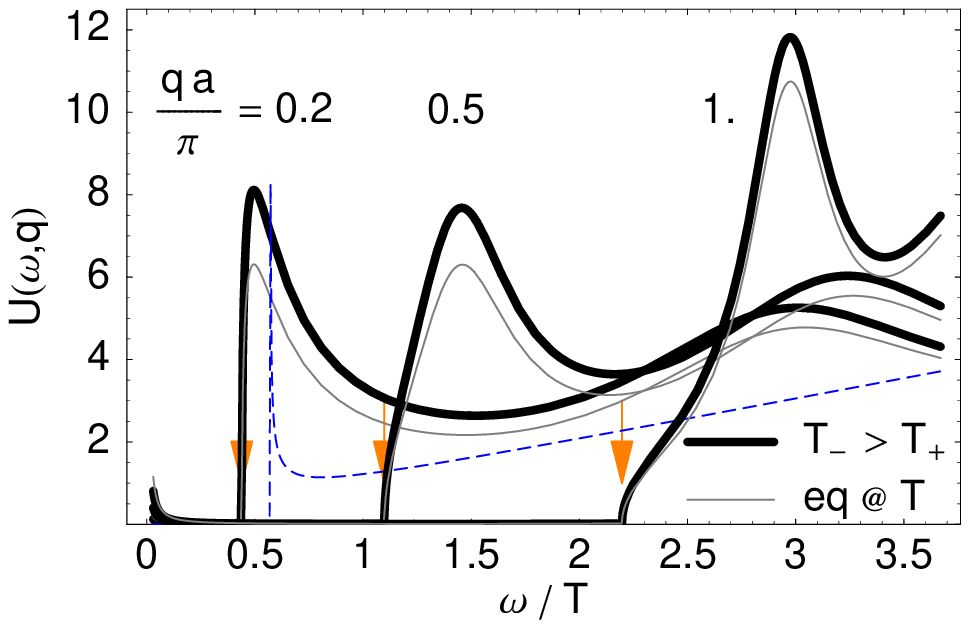}
\caption[]{Energy spectrum between a hot dielectric and a cold metallic
plate. (\emph{left panel}) The contours give the dimensionless spectrum 
$U( \omega, q )$ in steps of
$1$ from $0$ to $17$. Red lines: modes in a perfectly reflecting cavity.
Dotted green line: light cone $\omega = q$, orange line: polariton cone
in the dielectric $\omega = q / n$. The blue arrows mark the cuts shown
in the (\emph{right panel}): energy density vs.\ frequency $\omega$.
Thin lines: two different temperatures (same as left panel), thin gray
lines: global equilibrium at the average temperature $T = \frac12( T_+
+ T_-)$, dashed line: free-space spectrum at $T = 0$.
\\
Parameters (for a reference temperature $T = 300\,{\rm K}$): dielectric at $T_- = 390\,{\rm K}$ with index $n = 1.3$, 
metal at $T_+ = 210\,{\rm K}$ with impedance $\zeta( \omega )$ such that the 
skin depth at $\omega = T$ is $\approx 31\,{\rm nm}$. 
We calculate the impedance from a Drude conductivity with relaxation
time $\tau = 1.1 / T \approx 28\,{\rm fs}$. 
We have taken a relatively large distance $a = 1.1\,\lambda_T$ in order
to push the cavity resonances (red lines) into the thermal spectral range
($\lambda_T = 1/T \approx 7.6\,\mu{\rm m}$).
}
\label{fig:hot-n-cold}
\end{figure*}

To illustrate the general case, we use 
Eq.(\ref{eq:energy-spectrum-vs-KG}) and the
representation~(\ref{eq:6.5}) for $\hat D^K( \Omega; z, z' )$. In addition, we
integrate the energy density over the cavity volume $- a/2 < z < a/2$ in order
to reduce the number of relevant parameters.
The resulting spectrum $U( \omega, {\bf q} )$ of the energy per area
is dimensionless and plotted in the following as
a function of frequency $\omega$ and wave vector ${\bf q}$. 
We consider
for illustration purposes two complementary situations: (a) two dielectric
bodies with frequency-independent permittivity (index) $\varepsilon = n^2$
at zero temperature $T_\pm = 0$, the upper one moving at velocity $v$ along
the $x$-axis. Situation (b) is taken in mechanical equilibrium ($v = 0$)
at two different temperatures $T_+ \ne T_-$. One body is metallic, 
the other one dielectric as before. 

Fig.\ref{fig:sliding-dielectrics} illustrates the momentum distribution
$U( \omega, {\bf q} )$ of the energy spectrum in the ${\bf q}$ plane, 
at fixed frequency $\omega$. The parameters of the bodies are given in the
caption. By inspection of the formulas, 
we find that the surface sources have a spectral support inside the
`polariton cone' where the medium wave
vector $q_{z\varepsilon}$ is real
[see Eq.(\ref{eq:2.4})], i.e., $q \le n \omega$ (orange circle). 
If the dielectric is moving, the border of the polariton cone
is described by the equation
$q_{z\varepsilon}' = 0$ or explicitly
\begin{equation}
[ 1 - (nv)^2 ] q_x^2 + 2 (n^2 - 1) v \omega q_x + q_y^2 = (n^2 - v^2) \omega^2
\label{eq:elliptic-cone}
\end{equation}
For small enough velocity $v$, this describes an ellipse 
(left panel, red) that intersects the $q_x$ axis at
\begin{equation}
	q_{x1,2} = \omega \frac{ v \pm n }{ 1 \pm n v }
\label{eq:ellipse-points}
\end{equation}
Above the Cherenkov threshold, i.e., $v > 1/n$, Eq.(\ref{eq:elliptic-cone})
describes two 
hyperbolas (right panel, red line). It is interesting that the simple setting 
of a dielectric in fast motion creates a situation quite similar
to so-called hyperbolic or indefinite media. These have been studied
recently; they show similar dispersion relations in the bulk
and are approximately realized
in meta-materials with an anisotropic dielectric response \cite{Smith_2004, Smolyaninov_2010, Guo_2012, Biehs_2012}.

A setting with two temperatures is illustrated
in Fig.\ref{fig:hot-n-cold}: a hot dielectric facing a cold metal, both
at rest. Here, cylindrical symmetry holds
and the energy spectrum
$U( \omega, q )$ depends only on the modulus $q$ of the parallel wave vector.
In the $q\omega$-plane, one identifies the light and polariton cones 
(dotted green and orange), and the resonances of the planar cavity (red lines).
The latter are quite weak because the dielectric plate is a poor reflector.
The right panel in Fig.\ref{fig:hot-n-cold} shows broad peaks in the energy
density at these resonances, as well as sharper features just inside the 
polariton cone (arrows). The spectrum differs from a global equilibrium situation
(thin gray lines). This difference becomes small if $T_\pm$ are close,
as expected, but also for a highly conducting metal. The energy density is positive
everywhere because we did not subtract
the vacuum energy density (dotted blue line). The latter eventually dominates at
large frequencies.

\bigskip\

\acknowledgments

One of us (V.E.M.) acknowledges financial support by the European Science
Foundation (ESF) within the activity ``New Trends and Applications of the
Casimir Effect" (Exchange Grant 2847), and 
by the Deutsche Forschungsgemeinschaft (grant He-2849/4-1).
%
V.E.M. thanks Prof. M. Kardar and M. Kr\"uger for
fruitful discussions and P. Milonni, G. V. Dedkov and A. A. Kyasov for
comments on the manuscript. 
C.H. is indebted to G. Pieplow and H. Haakh for constructive
criticism.

%

\providecommand{\WileyBibTextsc}{}
\let\textsc\WileyBibTextsc
\providecommand{\othercit}{}
\providecommand{\jr}[1]{#1}
\providecommand{\etal}{~et~al.}

\end{document}